 \documentclass[twocolumn]{aa}
\usepackage{graphicx}
\usepackage{amssymb}
\begin{document}
   \title{Tracing baryons in the warm-hot intergalactic medium with broad Ly\,$\alpha$ absorption
   \thanks{Based on observations obtained with the NASA/ESA {\it Hubble Space Telescope},
    which is operated by the Association of Universities for Research in Astronomy, Inc.,
    under NASA contract NAS\,5-26555.}}


   \author{P. Richter,
          \inst{1}
          B.D. Savage,
          \inst{2}
	  K.R. Sembach,
          \inst{3}
	  \and
	  T.M. Tripp
          \inst{4}
          }

   \offprints{P. Richter\\
   \email{prichter@astro.uni-bonn.de}}

   \institute{Institut f\"ur Astrophysik und Extraterrestrische Forschung,
              Auf dem H\"ugel 71, 53121 Bonn, Germany
         \and
             Department of Astronomy, University of Wisconsin-Madison, 
	     475 N. Charter St., Madison, WI, 53706, USA
          \and
             Space Telescope Science Institute, 3700 San Martin Drive,
             Baltimore, MD 21218, USA
	    \and
	       Department of Astronomy, University of Massachusetts, 
	       Amherst, MA 01003, USA
            }

   \date{Received xxx; accepted xxx}

\abstract{
We discuss physical properties and baryonic content of broad 
Ly\,$\alpha$ absorbers (BLAs)
at low redshift. These absorption
systems, recently discovered in high-resolution,
high-signal to noise quasar absorption line spectra,
possibly trace the warm-hot intergalactic medium (WHIM)
in the temperature range between $10^5$ and $10^6$ K. The 
central idea is that in ionization equilibrium
WHIM filaments 
should contain a very small fraction of neutral gas
($f_{\rm H\,I}\sim 10^{-5}-10^{-6}$, typically), 
giving rise to weak intervening H\,{\sc i}
Ly\,$\alpha$ absorption. Due to the high temperature of the WHIM,
these Ly\,$\alpha$ absorbers must be thermally broadened to
Doppler parameters ($b$ values) $\geq 40$ km\,s$^{-1}$. 
It is expected, however, that also non-thermal line broadening 
processes, line blends, and noise features can mimic broad spectral features,
complicating the quantitative estimate of the baryon content of
the BLAs.
To extend previous BLA measurements
we have reanalyzed archival STIS data of the two quasars H\,1821+643 and
PG\,0953+415 and have identified 13 BLA
candidates along a total (unblocked) redshift path of $\Delta 
z=0.440$. Combining our measurements with previous results
for the lines of sight toward PG\,1259+593 and PG\,1116+215,
the resulting new BLA sample consists of 20 reliably detected 
systems as well as 29 additional tentative cases, implying
a BLA number density of $dN_{\rm BLA}/dz=22-53$. Eight BLAs
show associated absorption from H\,{\sc i} Ly\,$\beta$ 
and/or O\,{\sc vi}. The comparison between Ly\,$\alpha$, Ly\,$\beta$,
and O\,{\sc vi} line widths suggests that non-thermal broadening
and noise features substantially affect the observed BLA $b$ value distribution.
However, it remains unclear whether BLAs and O\,{\sc vi} absorbers
trace the same gas phase in the WHIM filaments.
We estimate that the contribution of BLAs to the 
baryon density at $z=0$ is $\Omega_{b}$(BLA)$\geq 0.0027\,h_{70}\,^{-1}$
for absorbers with log $[N($cm$^{-2})/b($km\,s$^{-1})] \gtrsim 11.3$.
This number indicates that WHIM broad Ly\,$\alpha$ absorbers 
contain a substantial fraction of the baryons in the
local Universe.
}
\titlerunning{Broad Ly\,$\alpha$ absorbers}

\maketitle
%

\section{Introduction}

Photoionized and collisionally ionized intergalactic gas 
most likely makes up for most of the baryonic matter in the local
Universe. While the diffuse photoionized intergalactic 
medium (IGM) that gives rise to the Lyman $\alpha$
forest accounts for $\sim 30$ percent of the 
baryons today (Penton, Stocke, \& Shull 2004), 
the shock-heated warm-hot intergalactic medium (WHIM)
at temperatures $T\sim 10^5-10^7$ K is expected to contribute
at a comparable level to the cosmological mass density of 
the baryons at $z=0$ (Cen \& Ostriker 1999; Dav\a'e et al.\,2001).
Gas and stars in galaxies, groups of galaxies, and galaxy 
clusters make up the rest of the baryonic mass (Fukugita 2003).

The WHIM is believed to emerge from intergalactic gas
that is shock-heated to high temperatures as the medium is 
collapsing under the action of gravity (Valageas, Schaeffer, \& Silk 2002).
Directly observing this gas phase is a particularly challenging task,
as the WHIM represents a low density ($n_{\rm H}\sim10^{-6}-10^{-4}$ cm$^{-3}$), 
high-temperature ($T\sim 10^5-10^7$ K) plasma, primarily made of 
protons, electrons, He$^+$, and He$^{++}$, together with traces of some 
highly ionized heavy elements. Diffuse emission from
this plasma is expected to have a very low surface brightness
and its detection awaits UV and X-ray
observatories more sensitive than currently available 
(see, e.g., Fang et al.\,2005; Kawahara et al.\,2005; Sembach et al.\,2005).
A promising approach to
study the WHIM is the search for absorption features from the 
WHIM in the far-ultraviolet (FUV) and 
in the X-ray regime. Five-times ionized 
oxygen (O\,{\sc vi}) is the most important high ion
to trace the WHIM at temperatures of
$T\sim 3\times 10^5$ K
in the FUV regime. Oxygen is a
relatively abundant element and the two available O\,{\sc vi}
transitions (located at $1031.9$ and $1037.6$ \AA) have
large oscillator strengths.
A number of detections
of intervening WHIM O\,{\sc vi} absorbers at $z<0.5$ 
have been reported in the literature 
(Tripp, Savage, \& Jenkins 2000; Oegerle et al.\,2000;
Tripp \& Savage 2000; Chen \& Prochaska 2000; Savage et al.\,2002;
Richter et al.\,2004, Sembach et al.\,2004; Savage et al.\,2005;
Tripp et al.\,2005; Danforth \& Shull 2005).
These measurements imply a number density of O\,{\sc vi}
absorbers per unit redshift of $dN_{\rm OVI}/dz \approx 17\pm 3$
for equivalent widths $W_{\lambda}\geq30$ m\AA\,(Danforth \& Shull 2005).
Assuming that 20 percent or less of the oxygen is present
in the form of O\,{\sc vi} ($f_{\rm O\,VI}\leq0.2$) 
and further assuming a mean oxygen abundance of $0.1$ solar, 
the measured number density of O\,{\sc vi} absorbers
corresponds to a cosmological mass density
of $\Omega_b$(O\,{\sc vi})$\geq0.0022$ $h_{70}\,^{-1}$.
For the interpretation of this value it has to be noted
that O\,{\sc vi} absorption mainly traces gas with
temperatures around $3 \times 10^5$ K, but not the
million-degree gas phase which probably contains 
the majority of the baryons in the WHIM.
Very recently, Savage et al.\,(2005) have reported the 
detection of Ne\,{\sc viii} in an absorption system
at $z\approx 0.2$ in the direction of the quasar HE\,0226$-$4110.
Ne\,{\sc viii} traces gas at $T\sim 7\times10^5$ K (in collisional
ionization equilibrium) and thus
is possibly suited to complement the O\,{\sc vi} measurements
of the WHIM in a higher temperature regime. 
However, as the cosmic abundance of Ne\,{\sc viii} is relatively low,
Ne\,{\sc viii} is not expected to be a particularly sensitive tracer of the
WHIM at the signal-to-noise levels achievable with current UV
spectrographs.
This is supported by the non-detections of intervening Ne\,{\sc viii}
in other high S/N STIS data (Richter et al.\,2004).
X-ray absorption
measurements are very important for studying the WHIM but are
currently limited in scope because of the small number of
available background sources and the relatively low spectral
resolution of current X-ray observatories (FWHM$\sim500$ to
$1000$ km\,s$^{-1}$). Weak WHIM X-ray absorption by O\,{\sc vii} and
O\,{\sc viii} has been
observed with the {\it Chandra} X-ray observatory towards
the quasar H\,1821+643 (Mathur, Weinberg, \& Chen 2003).
More recently, Nicastro et al.\,(2005) have reported the detection
of O\,{\sc vii} absorption in a WHIM filament in the {\it Chandra} spectrum
of the flaring blazar Mrk\,421.
In the future, more sensitive X-ray observatories such as
the {\it X-ray Evolving Universe Spectrometer} (XEUS) and
{\it Constellation-X} will be  of 
great importance to study the properties of the warm-hot
intergalactic gas over the entire temperature range in both
absorption and emission.

Next to high-ion absorption from oxygen and other metals,
recent observations with STIS (Richter et al.\,2004; Sembach
et al.\,2004) suggest that WHIM filaments can be detected
in Ly\,$\alpha$ absorption of {\it neutral} hydrogen. 
Although the vast majority of the hydrogen in the WHIM 
is ionized (by collisional processes and UV radiation), a 
tiny fraction ($f_{\rm H\,I}<10^{-5}$, typically) of neutral 
hydrogen should be present. Depending on the total
gas column density of a WHIM absorber and its 
temperature, weak H\,{\sc i} Ly\,$\alpha$ absorption
at column densities $12.5\leq$ log $N$(H\,{\sc i})$\leq 14.0$
may arise from WHIM filaments and could be used to
trace the ionized hydrogen component.
The Ly\,$\alpha$ absorption from WHIM filaments is 
expected to be very broad due to thermal line
broadening, resulting in large Doppler parameters
of $b>40$ km\,s$^{-1}$. Such lines are generally difficult 
to detect, as they are broad and shallow. High resolution,
high S/N FUV spectra of QSOs with smooth background continua
are required to successfully search for broad Ly\,$\alpha$
absorption in the low-redshift WHIM. The {\it Space Telescope
Imaging Spectrograph} (STIS) is 
the only instrument that has provided such data,
but due to the instrumental limitations of space-based
observatories, the number of QSO spectra adequate for searching 
for WHIM broad Ly\,$\alpha$ absorption (in the following abbreviated as `BLA')
is very limited.
So far, two sight lines observed with STIS 
towards the quasars PG\,1259+593 
($z_{\rm em}=0.478$) and PG\,1116+215 ($z_{\rm em}=0.176$)
have been carefully inspected for the presence of 
BLAs, and a number of candidates 
have been identified (Richter et al.\,2004; Sembach et al.\,2004).
These measurements imply a BLA
number density per unit redshift 
of $dN_{\rm BLA}/dz \approx 20-90$ for Doppler parameters
$b\geq40$ km\,s$^{-1}$ and restframe equivalent widths $W\geq30$ m\AA.
The large range for $dN_{\rm BLA}/dz$ partly is due to the uncertainty
about defining reliable selection criteria for
separating spurious cases from good broad Ly\,$\alpha$ candidates
(see discussions in Richter et al.\,2004 and Sembach et al.\,2004).
Transforming the number density $dN_{\rm BLA}/dz$ into a cosmological
baryonic mass density, one obtains $\Omega_b$(BLA)$\gtrsim 0.004\,h_{70}\,^{-1}$.
This limit is only about 10 percent of the total baryonic mass density
expected from the current cosmological models (see above), but
is above the limit derived for the 
intervening O\,{\sc vi} absorbers 
($\Omega_b$(O\,{\sc vi})$\geq 0.0022$ $h_{70}\,^{-1}$; 
see above).
Clearly, it is of great importance to extend the current sample 
of BLA candidates using recently obtained and archival STIS data
to more reliably deduce the baryonic content of the WHIM at low $z$.

In this paper we discuss the physical properties of the BLAs 
in the low-redshift Universe and review the observational constraints
that limit the detectability of these systems.
Moreover, we reanalyze the high S/N 
STIS spectra of the two QSOs H\,1821+643 ($z_{\rm em}=0.297$)
and PG\,0953+415 ($z_{\rm em}=0.239$) to search for WHIM
broad Ly\,$\alpha$ absorption along these two sight lines
(Sect.\,3). We combine our data with the results from
previous measurements for PG\,1259+593 and PG\,1166+215
and discuss physical properties and baryon content of the BLAs
in Sect.\,4. 
A summary of our study is given in 
Sect.\,5.

\section{Broad Ly\,$\alpha$ absorption}

\subsection{Motivation}

Ly\,$\alpha$ lines arising in the WHIM will be
thermally broadened, with $b$ values exceeding $\sim 40$ km\,s$^{-1}$.
Since the gas is ionized, only a very small fraction of the 
hydrogen will be in neutral form and the 
H\,{\sc i} column density is expected to be small.
Ly\,$\alpha$ absorption from the WHIM therefore 
should occur 
in the form of very broad, shallow absorption
lines. Their shape would differ significantly from
the shape of ordinary Ly\,$\alpha$ forest lines,
which are usually relatively narrow ($b\sim 25$ km\,s$^{-1}$; 
see Richter et al.\,2004).
For pure thermal broadening and assuming a collisional
ionization equilibrium, one can estimate the total hydrogen
content of each absorber directly from the measured
line width ($b$ value) and the H\,{\sc i} column
density (see Sect.\,2.2). However,
in addition to thermal line broadening, non-thermal
broadening mechanisms exist, such as 
the Hubble flow or gas flows and turbulence 
within the cosmic web filaments. Also unresolved
line blends can produce absorption features 
that are broad and shallow.
For the analysis of interstellar and intergalactic
absorption lines it is therefore common to assume that
the measured $b$ value of an absorption
line is composed of a thermal component, $b_{\rm th}$,
and a non-thermal component, $b_{\rm non-th}$, so 
that :
\begin{equation}
b=\sqrt{b_{\rm th}\,^2+b_{\rm non-th}\,^2}.
\end{equation}
Current FUV absorption line data typically have 
limited spectral resolution 
and moderate signal-to-noise ratios (S/N). This
means that some of the apparently broad, shallow 
absorber candidates in these spectra
may be produced by noise features, blends of multiple
narrow absorption lines, and continuum undulations
rather than by WHIM gas.
To reliably estimate the baryon content of the
WHIM from broad Ly\,$\alpha$ absorption it is therefore necessary
to assess the effects of non-thermal broadening 
processes and data quality issues. In the following
sections we discuss these various aspects 
in more detail to identify difficulties for 
the interpretation of broad spectral features.

\subsection{Thermal line broadening}

\begin{figure}[t!]
\resizebox{1.0\hsize}{!}{\includegraphics{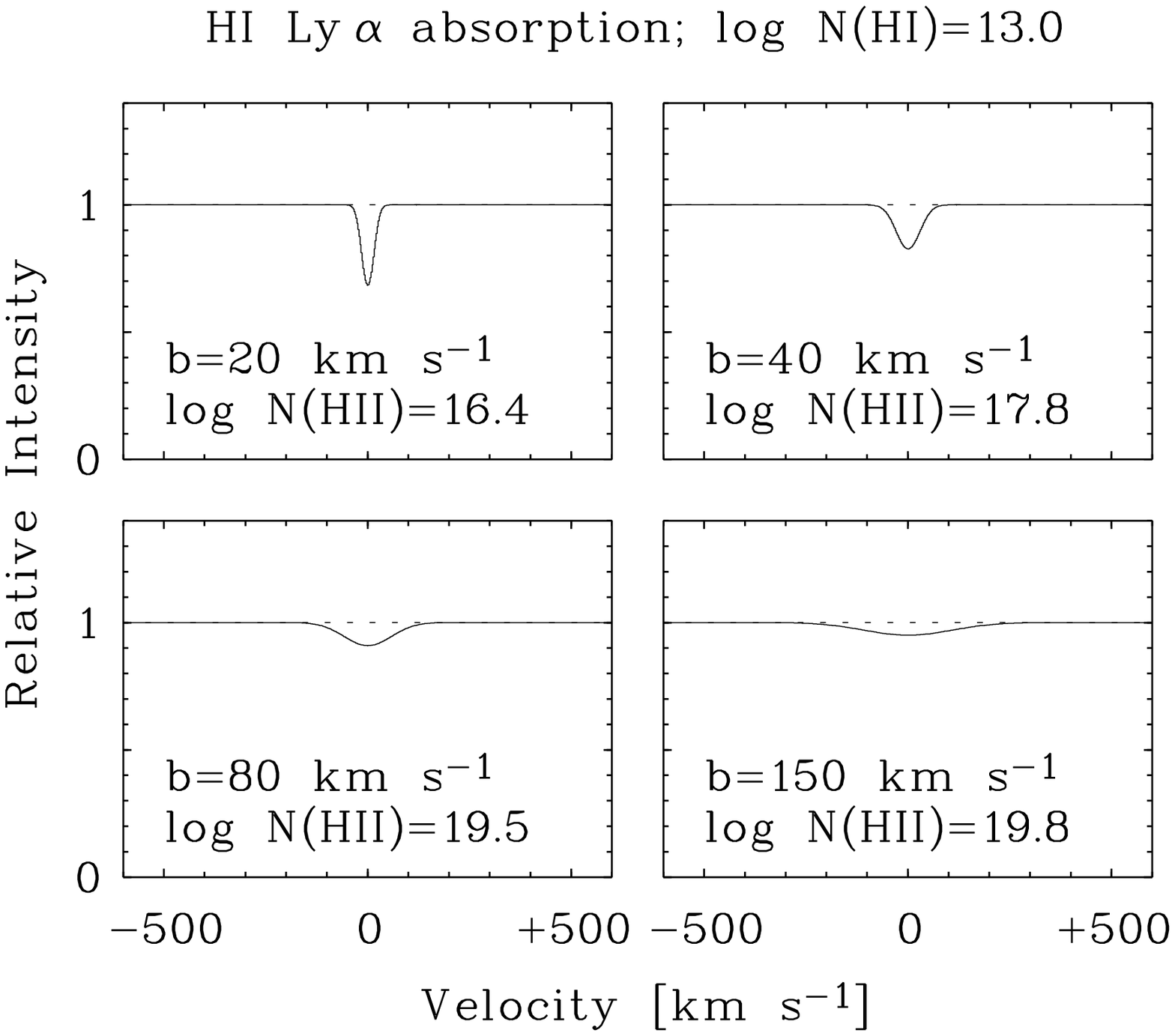}}
\caption[]{Theoretical Voigt line profiles for H\,{\sc i} Ly\,$\alpha$ 
absorption with a constant H\,{\sc i} column density of log $N$(H\,{\sc i})$=13.0$ 
and $b$ values of $20, 40, 80$, and $160$ km\,s$^{-1}$ are shown. For pure thermal broadening
and in collisional ionization equilibrium, these lines would trace gas with 
ionized gas column densities of log $N$(H\,{\sc ii})$=16.4, 17.8, 19.5$, and $19.8$,
respectively.}
\end{figure}

If the width of an absorption line is governed
solely by thermal broadening (i.e., $b_{\rm non-th}=0$),
the measured $b$ value of the line
serves as a direct measure for the
temperature of the gas:
\begin{equation}
b =  \sqrt \frac{2kT}{m} \approx 0.129 \sqrt \frac{T}{A}\, \rm{km\,s}^{-1},
\end{equation}
where $k$ is the Boltzmann constant, $m$ is the particle mass, 
and $A$ is the atomic weight of the element.
We demonstrate the effect of thermal line
broadening of the Ly\,$\alpha$ absorption in
Fig.\,1:
we model the expected 
shape (Voigt profile) of an H\,{\sc i} Ly\,$\alpha$ absorption
line with a constant H\,{\sc i} column density
of log $N$(H\,{\sc i})$=13.0$
for Doppler parameters $b=20, 40, 80$, and $150$ km\,s$^{-1}$.
From equation (2) follows that $T\approx 60b^2$ for
pure thermal broadening.
The $b$ values of $20, 40, 80$, and $150$ km\,s$^{-1}$ 
given in Fig.\,1 then correspond to temperatures of
log $T\approx 4.4, 5.0, 5.6$, and $6.1$.
Clearly, it is much easier to detect Ly\,$\alpha$
absorption from gas with $T\approx 10^4$ K
than  Ly\,$\alpha$
absorption from gas with $T\approx 10^6$ K.
The search for broad Ly\,$\alpha$ absorption
from the WHIM at temperatures $T=10^5-10^6$ K
therefore requires spectral data with good S/N 
and a smooth background continuum.

If collisional ionization equilibrium (CIE) applies, one
can directly calculate the hydrogen ionization fraction
log $f_{\rm H}=$log ((H$^0+$H$^+)/$H$^0)\approx$
log (H$^+/$H$^0$)
as a function of the gas temperature 
(Sutherland \& Dopita 1993; Richter et al.\,2004) :
\begin{equation}
{\rm log}\,f_{\rm H}(T)\approx -13.9 + 5.4\,{\rm log}\,T - 0.33\,
({\rm log}\,T)^2.
\end{equation}
Together with the measured H\,{\sc i} column density, $N$(H\,{\sc i}), one
can determine the total hydrogen column density
via $N$(H$)=f_{\rm H}\,N$(H\,{\sc i}).
Under these assumptions, the chosen sequence of $b$ values
in Fig.\,1 not only is a sequence of 
increasing temperature of the absorbing gas,
but also a sequence of 
increasing ionized hydrogen column density 
(log $N$(H\,{\sc ii}$)\approx 16.4, 17.8, 19.5$, and $19.8$).
The correlation between higher gas content and weaker
H\,{\sc i} absorption has important observational consequences
if the lines are thermally broadened:
{\it the broader and shallower a WHIM
H\,{\sc i} Ly\,$\alpha$ line is, the
more baryonic material it traces.} This 
means that it will be particularly difficult to detect
those WHIM Ly\,$\alpha$ absorbers that have
the largest baryon content, as many of these
features may 
fall below the detection limit.
Note that for low gas densities ($n_{\rm H}<10^{-5}$ cm$^{-3}$)
photoionization by the UV background contributes to
the ionization of the WHIM (see Sect.\,2.6). Therefore,
the total hydrogen column density derived from equation (3)
under the assumption of CIE 
has to be regarded as a lower limit for $N$(H) in BLAs.

\subsection{Non-thermal line broadening}

\begin{figure*}[t!]
\resizebox{1.0\hsize}{!}{\includegraphics{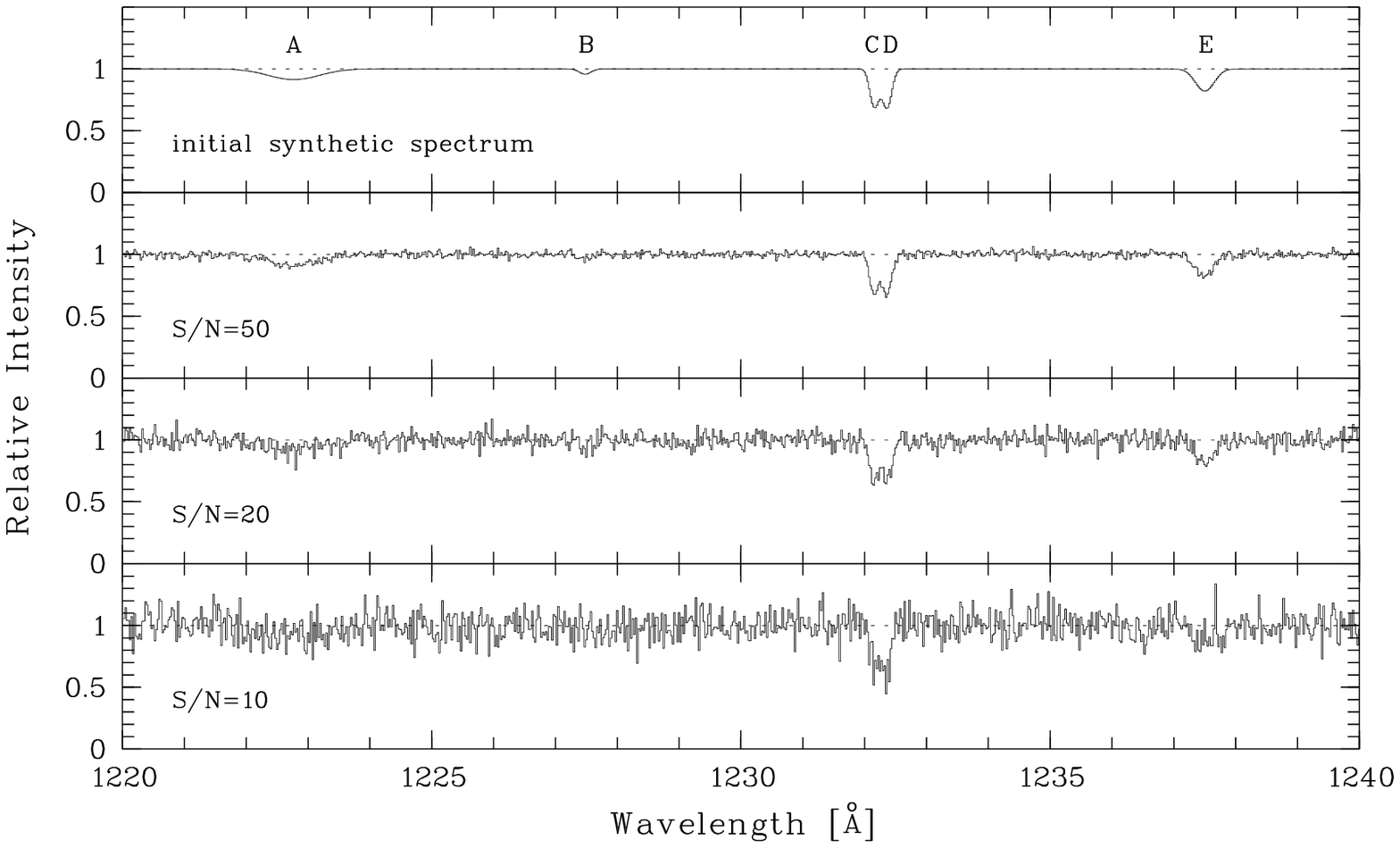}}
\caption[]{Artificial spectra of Ly\,$\alpha$ absorption in the 
wavelength range between $1220$ and $1240$ \AA\, are shown. The upper 
panel shows the initial synthetic spectrum (without noise) with
five Ly\,$\alpha$ absorbers A, B, C, D, and E having
log $N$(H\,{\sc i})$=13.23, 12.28, 13.07, 13.10, 13.12$ and
$b$ values of $141, 32, 24, 25, 51$ km\,s$^{-1}$, respectively.
The lower three panels show the same spectrum with Poisson noise 
added at S/N levels of 50, 20 and 10 per 7 km\,s$^{-1}$ wide pixel.
For a S/N$\leq20$ the broad Ly\,$\alpha$ absorber A almost 
vanishes in the noise, and the two narrow absorbers C and D
start to smear together to one broad spectral feature. At a S/N of 10,
the broad Ly\,$\alpha$ absorber E is affected by a noise
peak and thus looks asymmetric. These spectra demonstrate
the difficulty to reliably identify intrinsically broad 
Ly\,$\alpha$ absorbers in data with limited S/N.
}
\end{figure*}

The network of WHIM filaments represents cosmic structures at scales
of a several Mpc (e.g., Cen \& Ostriker 1999; Dav\'e et al.\,2001), while
the thickness of an individual filament is expected to be on the
order of several hundred kpc (Valageas, Schaeffer, \& Silk 2003).
Line broadening due to peculiar
velocities and turbulence in the 
gas therefore must be considered
when it comes to the interpretation of lines widths
of WHIM absorbers. Clearly, at such large scales
the gas is expected to have density and velocity
substructure that will leave its imprint in the
observed H\,{\sc i} optical depth distribution
(Kawahara et al.\,2005).
These effects would result in
non-Gaussian shapes of broad H\,{\sc i} lines,
which could be identified in absorption line
data at sufficient spectral resolution and S/N.
Broad H\,{\sc i} lines that are asymmetric therefore
should be interpreted with particular care, as their
width may not represent a good measure of the
temperature of the gas. 

As spectral resolution and signal-to-noise (S/N) in FUV
data are limited, apparently broad Ly\,$\alpha$
absorption may be caused by individual 
unresolved velocity components. 
In particular
the measurements of line widths from intermediate and low spectral
resolution spectra (FWHM$ \geq 20$ km\,s$^{-1}$) often turn out
to be imprecise due to the presence of subcomponents that
are not, or only barely resolved in the data.
The resolution of the STIS E140M spectra used in
this and the previous two studies is high, and with a FWHM
of $\sim 7$ km\,s$^{-1}$,
much smaller than the typical widths of the
lines under investigation ($b\geq 40$ km\,s$^{-1}$ or
FWHM$\geq 66.5$ km\,s$^{-1}$). At this resolution, most
of the existing subcomponent structure in intervening
absorbers should be resolved. Multiple absorption components
that are separated by $\sim 20-40$ km\,s$^{-1}$ will in most cases produce
an absorption structure that looks asymmetric and non-Gaussian.
For uniformity, in this
paper we concentrate on STIS E140M spectra, but we note that BLA
candidates have also been identified in lower-resolution observations
with the STIS first-order gratings (e.g., Bowen, Pettini, \& Blades
2002).

If the WHIM gas
participates in the overall Hubble-flow, Hubble broadening
will contribute with $\sim 7$ km\,s$^{-1}$ per
hundred kpc thickness of the absorber to the total
line width. However, if the thickness of the absorbing WHIM structures
is on the order of a few hundred kpc (Valageas, Schaeffer, \& Silk 2002),
one expects that $b_{\rm Hubble}\leq 30$ km\,s$^{-1}$, typically.
Moreover, as the WHIM is produced by collapsed gaseous structures
that undergo gravitational shock-heating, one may argue 
that the resulting WHIM filaments primarily have decoupled from
the Hubble flow during the collapse phase and that the
gas dynamics is driven primarily by internal processes
(this should be valid at least
for the main direction of the collapse). 
Hubble broadening therefore is not expected to dominate
the widths of WHIM spectral lines.
For those BLAs in which
associated O\,{\sc vi} absorption is observed,
one can evaluate the possible presence of non-thermal
line broadening by comparing the $b$ value of
H\,{\sc i} and O\,{\sc vi}. For pure thermal
broadening the O\,{\sc vi} $b$ value is expected
to be four times smaller than that of H\,{\sc i} owing to
the 16 times larger mass of oxygen (see equation 2).
Non-thermal contributions to $b$ may be significant
for BLA/O\,{\sc vi} absorber pairs that have
$b_{\rm H\,I}<4\,b_{\rm O\,VI}$.
However, only a small fraction of the BLAs
show associated O\,{\sc vi} absorption, so that
this consistency check on the gas temperature 
can be applied only to a few cases.
In addition, significant velocity offsets 
between H\,{\sc i} and O\,{\sc vi}
as well possible
multi-component structure in the O\,{\sc vi} are
observed in some cases.
It thus remains unclear whether the H\,{\sc i}
and the O\,{\sc vi} absorption always trace
the same gas phase, and whether the comparison
between O\,{\sc vi} and H\,{\sc i} line widths
provides reliable information about non-thermal
gas motions.

\subsection{Noise}

FUV spectral data of extragalactic background sources 
obtained with STIS typically
have S/N$<30$ per $\sim 7$ km\,s$^{-1}$ 
wide resolution element. The presence
of noise clearly hampers the reliable detection of
WHIM BLA candidates. 
The S/N in the data will determine the detection limit
of broad features and so it is clear that
only the highest H\,{\sc i} column density WHIM filaments can be detected with
current FUV data. In contrast to the O\,{\sc vi} absorbers, the
detection limit for broad Ly\,$\alpha$ absorption cannot easily
be quantified in terms of a minimum equivalent width, as both
the depth {\it and} the width of a line limit its detectability.
For instance, a relatively narrow H\,{\sc i} Ly\,$\alpha$
line with $b=40$ km\,s$^{-1}$ and log $N=13.0$ (see Fig.\,1,
upper right panel) will have an equivalent width of
$\sim 50$ m\AA\,and can easily be detected with STIS data
at a S/N$\geq15$. At this S/N, however, a very broad line with
$b=150$ km\,s$^{-1}$ and log $N=13.0$ (Fig.\,1, lower right panel)
would not be detectable, although the equivalent width would be the
same. From the analysis of STIS data we find that
BLA candidates can be identified 
if the 
following condition is fulfilled:
\begin{equation}
\frac{N{\rm (H\,I)}}{b{\rm (H\,I)}}
\gtrsim \frac{3\times 10^{12}}{\rm (S/N)}\,\,{\rm cm}^{-2}\,({\rm km\,s}^{-1})^{-1}.
\end{equation}
In this equation, S/N is the local signal-to-noise ratio per
resolution element.
The S/N ratio may vary significantly
within a spectrum due to large-scale flux variations in the
QSO spectrum and the wavelength-dependent
sensitivity of the detector. The
average S/N in our STIS data is $\sim 15$, so that
the typical detection limit for detecting broad Ly\,$\alpha$
absorbers in these data is log $(N/b)_{\rm min} \gtrsim 11.3$.

Noise features will also influence
the shape of spectral features and thus will 
complicate the search for WHIM BLA candidates.
Therefore, we have studied the influence of noise features 
(Poisson noise) 
in a set of artificial spectra with various S/N levels.
The FWHM in these spectra is
similar to those in the STIS data and the pixel size 
corresponds to one resolution element ($7$ km\,s$^{-1}$).
An example spectrum is shown in Fig.\,2. The outcome 
of our noise analysis of these mock spectra is the following:
(1) noise can create structures in the continuum that
could be mis-identified as intervening broad 
Ly\,$\alpha$ absorption, (2) noise can create broad
absorption features out of multiple, individual velocity components
that are separated only by $\sim 20-50$ km\,s$^{-1}$, (3)
noise can substantially modify the shape of an
existing broad absorber, so that Gaussian-shaped
absorption lines become asymmetric, and vice versa.
The binning of several pixels can
help to check on the validity of an absorption feature,
but we find that at S/N$<20$ more than $50$ percent 
of all apparent broad absorption features are 
caused by noise rather than by intrinisically broad lines.
Next to Poisson noise, other systematic noise sources
can be important for the interpretation of spectral
features in FUV absorption line data. For the STIS
instrument, possible order-edge artifacts and features 
caused by the obscuration by the field 
electrode of the MAMA detector (the ``repeller wire'';
see Woodgate et al.\,1998) have to be considered.
The limited S/N in the currently existing FUV data 
introduces the largest uncertainty for the 
proper identification and analysis of 
BLA candidate lines. Note, however, that the broad
lines chosen for this study currently represent
the best ones available for the search of BLAs. 

\subsection{Continuum undulations}

Due to the large width and the shallowness of the BLAs,
one may argue that part of
these features are not produced by
intervening absorption but rather are caused caused by small-scale
undulations in the overlaying quasar continuum.
One simple way to test for the presence of continuum
undulations in the STIS data
is to analyze the smoothness of the quasar
continuum shortwards and longwards
of the QSO Ly\,$\alpha$ emission.
If some of the features identified
as broad Ly\,$\alpha$ absorbers are instead
produced by continuum undulations, one
should expect that such undulations are
distributed all over the spectrum, whereas
true intergalactic BLAs
can occur only in the wavelength range between
$\lambda_{\rm BLA,min}=1215.67$ \AA\, and
$\lambda_{\rm BLA,max}=(1+z_{\rm QSO}) \times 1215.67$ \AA.
However, in none of the analyzed spectra do broad
features (if significant with respect to the 
ambient noise) occur at wavelengths $\lambda_{\rm BLA,max}>
(1+z_{\rm QSO}) \times 1215.67$ \AA. This implies that
continuum undulations do not interfere with
our BLA search.

Continuum undulations also may be caused by 
weak emission lines from the QSO. Possible
candidate lines for the range 
$\lambda_{\rm BLA,max}<(1+z_{\rm QSO}) \times 1215.67$ \AA\,
are, for instance, H\,{\sc i} Ly\,$\beta\,\lambda 1025.7$,
O\,{\sc vi}\,$\lambda 1031.9$, C\,{\sc iii}\,$\lambda 977.2$.
However, none of the BLA candidates considered in this
paper lie close to the wavelengths at which QSO
emission in these (and other) lines would be expected.

\subsection{Deviations from CIE}

The determination of $f_{\rm H}$ and $N$(H\,{\sc ii})
using equation (3) is valid only under the assumption
that the gas is in collisional ionization
equilibrium, which may not be the case in general.
In fact, it is expected that for low-density 
gas ($n_{\rm H}<10^{-5}$ cm$^{-3}$) photoionization
from the UV background becomes important, so that
the ionization fraction (as defined in Sect.\,2.2)
in low-density BLAs should be higher compared to what is
estimated for pure collisional ionization.
To investigate the influence of photoionization we
have analyzed the output of a WHIM simulation presented
by Fang \& Bryan (2001). This simulation includes a combined
(time equilibrium) collisional ionization and photoionization model
(hereafter referred to as CIE+PH),
based on ionization fractions from Mazzotta et al.\,(1998) and
the CLOUDY photoionization code (Ferland et al.\,1998).
The WHIM simulation predicts that for 
$T=10^5-10^7$ K and $n_{\rm H}=10^{-6}-10^{-4}$ cm$^{-3}$
the CIE+PH ionization fraction is (on average) $\sim 50$ percent
higher than for CIE alone. If we restrict the density range
to $n_{\rm H}>10^{-5}$ cm$^{-3}$, the CIE+PH ionization fraction
typically lies $\sim 15$ percent above the CIE estimate.
This shows that photoionization is important for the BLAs.
However, since we do not have information about the actual hydrogen 
volume densities in these absorbers, we only can notice that with the 
CIE assumption we probably underestimate the ionized hydrogen column
density (and thus the baryon content)
of BLAs by $\sim 15-50$ percent. To more precisely pinpoint
the role of photoionization in BLAs a more detailed analysis of
broad Ly\,$\alpha$ absorption in cosmological
simulations is required. This will be presented in a future
paper (Richter et al.\,2005, in preparation).

\begin{table*}[t!]
\caption[]{Log of STIS observations}
\begin{tabular}{lrrcclcl}
\hline
Sight line & $l$ & $b$ & $V$ & $z_{\rm em}$ & Grating & Exposure time & Broad Ly\,$\alpha$ analysis \\
& & & [mag] & & & [ks] & \\
\hline
H\,1821+643  & 94.00  & $+27.42$ & 14.24 & 0.297 & E140M & 25.5 & this study \\
PG\,0953+415 & 179.79 & $+51.71$ & 15.32 & 0.239 & E140M & 24.5 & this study \\
PG\,1116+215 & 223.36 & $+68.21$ & 14.80 & 0.177 & E140M/E230M & 35.5 & Sembach et al.\,(2004) \\
PG\,1259+593 & 120.56 & $+58.05$ & 15.84 & 0.472 & E140M & 95.5 & Richter et al.\,(2004) \\
\hline
\end{tabular}
\end{table*}

\subsection{The baryonic content of broad Ly\,$\alpha$ absorbers}

The cosmological mass density of the WHIM broad Ly\,$\alpha$ absorbers
in terms of the current critical density $\rho_{\rm c}$ can
be estimated by
\begin{equation}
\Omega_b{\rm (BLA)}=\frac{\mu\,m_{\rm H}\,H_0}
{\rho_{\rm c}\,c}\,\sum_{ij}\,f_{{\rm H},ij}\,N({\rm H\,I})_{ij}\,
\Big / \sum_{j}\Delta X_j,
\end{equation}
with $\mu=1.3$, $m_{\rm H}=1.673 \times 10^{-27}$ kg, $H_0=
70$ km\,s$^{-1}$\,Mpc$^{-1}$, and $\rho_{\rm c}=3H_0\,^2/8 \pi G$.
The index $i$ denotes an individual broad Ly\,$\alpha$ system
along a line of sight $j$. Each measured absorption
system $i$ is characterized
by its neutral hydrogen column density, $N$(H\,{\sc i})$_{ij}$ and
ionization fraction, $f_{{\rm H},ij}$ (see equation 3).
Each line of sight $j$ has a
characteristic redshift range $\Delta z=z_{\rm max}-z_{\rm min}$
in which broad Ly\,$\alpha$ absorption may be detected.
The comoving path length $\Delta X_j$ 
available for the detection broad Ly\,$\alpha$
systems then is given by:
\begin{footnotesize}
\begin{equation}
\Delta X=0.5\,\{[(1+z_{\rm max}-\Delta z_{\rm B}/2)^2-1]-   
[(1+z_{\rm min}+\Delta z_{\rm B}/2)^2-1]\},
\end{equation}
\end{footnotesize}
where we refer to a cosmology with $q_0=0$.
Due to the presence of Ly\,$\alpha$ forest lines
and interstellar absorption lines a blocking correction $\Delta z_{\rm B}$
for each sight lines has to be applied.

\subsection{Summary: interpretation of H\,{\sc i} line widths}

As there currently are only very few techniques that can be used to
constrain the baryon budget of the WHIM by direct observations,
the analysis of broad Ly\,$\alpha$ absorbers in low-$z$ QSO absorption
line spectra is of fundamental importance for our understanding
of the distribution of baryonic matter in the local Universe.
Every WHIM filament that contains sufficient amounts of
neutral hydrogen will produce a broad Ly\,$\alpha$
absorption line, but, vice versa, not every broad 
feature observed in FUV data is caused by the WHIM. Moreover,
the quantitative estimate of $N$(H\,{\sc ii}) from
$N$(H\,{\sc i}) and $b$(H\,{\sc i}) is
hampered by the various alternative broadening mechanisms
discussed above. Therefore it is clear that the
interpretation of broad absorption features in FUV spectra
and the determination of $\Omega_b$ from BLA
candidates is afflicted with systematic uncertainties.
This should be kept in mind for the second part of the
paper, where we analyze BLA candidates along
four lines of sight and estimate their baryon content.

\section{STIS observations of low-$z$ broad Ly\,$\alpha$ absorbers}

\subsection{Overview}

STIS data with medium- to high spectral resolution and good S/N 
are available only for a very limited
number of QSO sight lines, as it requires a substantial
amount of observing time to observe objects
with $V$ magnitudes $>15$ at high spectral resolution.
In this study, we analyze broad Ly\,$\alpha$ absorbers
along the two lines of sight towards H\,1821+643 and PG\,0953+415
and combine these data with previously published measurements
for PG\,1259+593 (Richter et al.\,2004) and PG\,1116+215
(Sembach et al.\,2004). Information about the line-of-sight
properties is summarized in Table 1.
All four spectra were obtained between 1998 and 2002 
using the medium resolution
FUV echelle mode (E140M), which provides a spectral
resolution of $\lambda/ \Delta \lambda \approx 46,000$,
corresponding to FWHM$\approx 7$ km\,s$^{-1}$. 
The integration time for each spectrum varies between
$\sim 25$ and $\sim 95$ ks (see Table 1). More detailed information 
about the observations and primary data reduction of these
spectra can be found in Tripp, Savage, \& Jenkins (2000), 
Tripp \& Savage (2000), Richter et al.\,(2004), and
Sembach et al.\,(2004, 2005a).
Note that the STIS data sets for H\,1821+643 and 
PG\,0953+415 have been extensively used to study
the distribution and properties of intervening
O\,{\sc vi} absorbers at low redshifts
(Tripp, Savage, \& Jenkins 2000; Tripp \& Savage 2000;
Tripp et al.\,2001; Savage et al.\,2002), as well as to investigate the
statistical and physical properties of the low$-z$
Ly\,$\alpha$ forest (Dav\a'e \& Tripp 2001). In the
latter study, the authors compared the 
properties of the Ly\,$\alpha$ forest in these
two spectra with those of mock QSO spectra from
hydrodynamical simulations. For the analysis,
Dav\a'e \& Tripp used an automated continuum
fitting routine together with an automated Voigt
profile fitting algorithm (AutoVP; see Dav\a'e et al.\,1997)
to efficiently handle the large number of lines
in their data. With this method, they find
89 Ly\,$\alpha$ forest absorbers with
$b<40$ km\,s$^{-1}$
along a total redshift path of $\Delta z=0.335$
(Dav\a'e \& Tripp 2001; their Fig.\,6),
but apparently only two systems with $b\geq40$ km\,s$^{-1}$.
However, our experience with automated line fitting
routines is that broad and shallow spectral features
often are not identified by these routines, or that
they are inappropriately fitted, e.g., by multiple,
narrow lines. A careful re-analysis of these data
focussed on the search of broad and shallow absorption
features is justified.

\subsection{Data analysis}

\begin{figure*}[t!]
\resizebox{1.0\hsize}{!}{\includegraphics{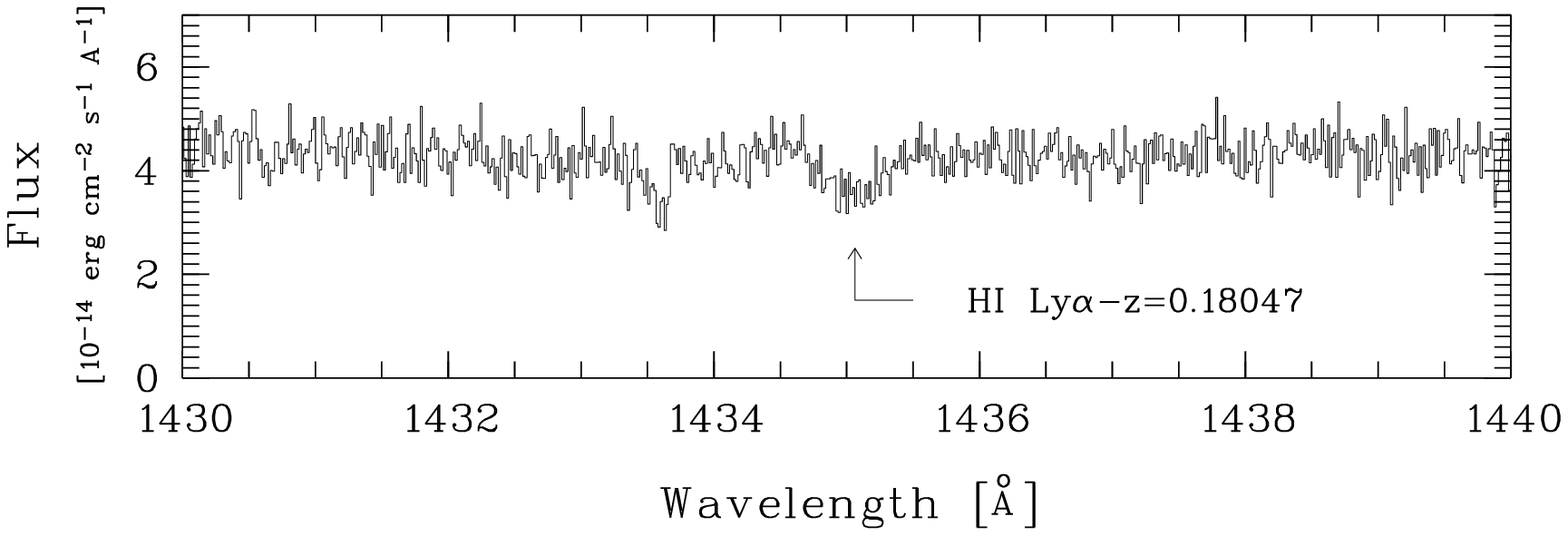}}
\caption[]{The well-detected broad Ly\,$\alpha$ absorber 
at $z=0.18047$ in the STIS spectrum of H\,1821+643 
is shown. This system has an H\,{\sc i} column density of log $N$(H\,{\sc i})$=13.21$
and a $b$ value of $\sim 56$ km\,s$^{-1}$ (see also Table 2). The shape of this BLA
differs significantly from the shape of the narrow $z=0.17924$ Ly\,$\alpha$ 
forest absorption near $1433.5$ \AA.}
\end{figure*}

For the data analysis of the STIS spectra of
H\,1821+643 and PG\,0953+415 we follow the
procedures from previous spectral anlyses
of STIS data for PG\,1259+593 (Richter et al.\,2004)
and PG\,1116+215 (Sembach et al.\,2004).
From a careful by-eye inspection of the STIS data
(put onto a heliocentric velocity scale)
we first have created a catalog of broad spectral
features that represent candidates for intervening
broad H\,{\sc i} Ly\,$\alpha$ absorption ($\lambda_0=1215.67$
\AA; oscillator strength $f=0.460$; Morton 2003).
As we do not want to include absorbers that
possibly are associated with the background
quasar itself, we have considered only those lines
that are more than $5000$ km\,s$^{-1}$ away
from the quasar.
We have omitted from our candidate list those absorption
features that exhibit clear evidence for sub-component
structure, line blending, and irregularities due
to continuum undulations. Moreover, we do not 
consider broad absorption features that are part
of complex multicomponent absorption systems, as the
width of these absorbers is particularly difficult
to determine. We therefore ignore hot gas that is
situated in multiphase absorbers that may
preferentially be related to gas in the vicinity 
of galaxies and galaxy groups. This introduces 
a selection bias in our BLA sample.
Note that recent observations of Ne\,{\sc viii} and O\,{\sc vi}
in a multiphase absorber at $z\approx0.2$ toward 
HE\,0226$-$4110 suggest that such systems may contain
significant amounts of hot gas (Savage et al.\,2005).
It is important to emphasize that our selection procedure
is based solely on the visual inspection of absorption
features in the spectrum. This method thus involves a certain
degree of subjectiveness for the identification of BLA
candidates, in particular if it comes to the evaluation
of noise features, line asymmetries, and continuum undulations.
As outlined earlier, automated fitting procedures
currently do not represent a reliable alternative to assess the
significance of broad spectral features. Therefore, we here
rely on selection criteria based on our experience with 
previous absorption line data. Note that the BLA candidates
presented in this paper have been identified and 
evaluated independently by the various
authors of this paper. 

After having selected a
sample of BLA candidate lines we have cross-checked whether
these features could be higher Lyman lines or metal
lines belonging to other intergalactic absorption systems,
and modified the list accordingly. We then checked 
for associated broad Ly\,$\beta$ and O\,{\sc vi} absorption
for the BLA candidates. BLAs that show associated
Ly\,$\beta$ absorption with $b_{\rm \,Ly\,\beta}<40$ km\,s$^{-1}$
have been excluded from our sample (see Sect.\,4.2.1).
For the final BLA candidate sample, the lines
have been continuum-normalized.
The continuum fitting was done for each line individually,
using low-order polynomials that fit the continuum
around a line in range from approximately $-1000$ to $+1000$
km\,s$^{-1}$. The base points for the polynomial fit
were obtained from a binned (10 pixels, typically) version
of the spectrum to smooth out the noise in the continuum
flux and thus to increase the reliability of the continuum fit.
The normalized broad Ly\,$\alpha$ candidate lines then
were fitted with Voigt profiles convolved with a
Gaussian line spread function that corresponds to the
spectral resolution of the STIS instrument (FWHM$\approx7$
km\,s$^{-1}$). As the lines under analysis generally
have FWHM$>50$ km\,s$^{-1}$, possible small deviations
of the STIS line spread function from a single-component Gaussian
have no measurable influence on the
outcome of the profile fit. For the fitting procedure
we have used the Voigt profile/ $\chi ^2$ minimization
routine FITLYMAN installed in the ESO-MIDAS software
package (Fontana \& Ballester 1996), which
delivers line centroids, redshifts, equivalent widths, logarithmic
column densities, $b$ values, and their corresponding
errors. The total uncertainty for $N$ and $b$
have been calculated by quadratically adding the error estimates
from the fitting process to those from the continuum
placement. Due to the fact
the S/N varies over the spectrum, the detection
limit for BLAs, log $(N/b)_{\rm min}$, as defined in Sect.\,2.4,
also varies across the spectrum.
Many of the BLA candidates in our sample have a low S/N in their
profile. In some cases, there is weak (but not
compelling) evidence for asymmetries and
unresolved component structure. 
Although these systems
are all clearly detected with $W_{\rm r}/ \sigma >4.4$ except
for one line with $W_{\rm r}/ \sigma =2.6$, 
it is probable that some
of them are not related to thermally broadened H\,{\sc i}
but are caused by blends of several subcomponents and
noise features. As the S/N in our data is not good enough to
reliably evaluate the true component structure in these
broad lines, we regard these absorbers as tentative BLA 
detections. Throughout the following we label these
cases accordingly.
Measured column densities, $b$ values, and
other quantities for each absorber in our final BLA
candidate sample are listed in Tables 2 and 3. In Table 4
we list estimates for the temperature, ionization fraction,
and total hydrogen column density for each system (see 
Sect.\,4.3). Statistical information for each sight line
is presented in Table 5.

\subsection{H\,1821+643}

\begin{figure}[t!]
\resizebox{0.95\hsize}{!}{\includegraphics{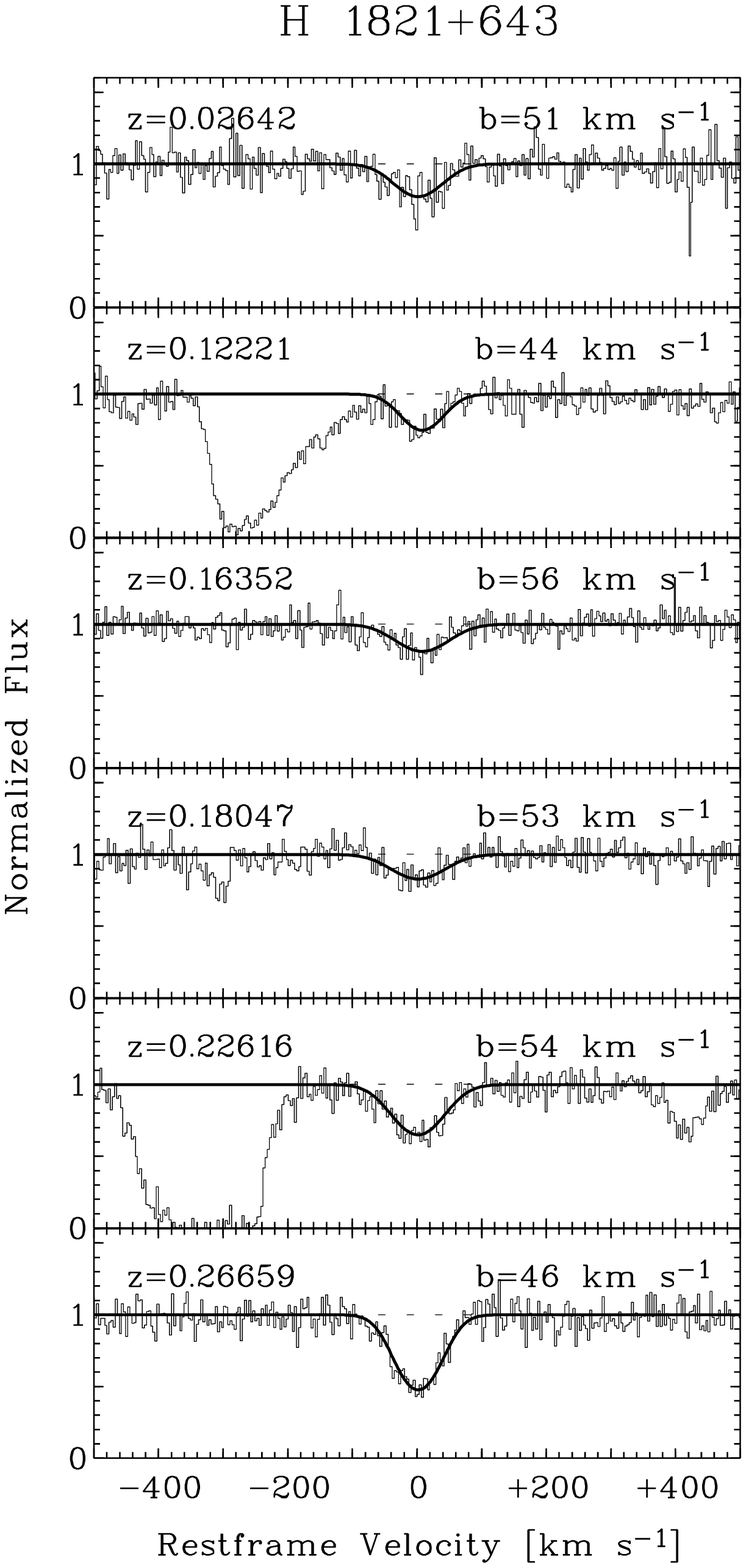}}
\caption[]{Broad Ly\,$\alpha$ absorbers towards H\,1821+643, plotted
on a restframe velocity scale. 
Voigt-profiles fits are indicated with the solid line. Other absorption
features are related to the Ly\,$\alpha$ forest and metal lines.}
\end{figure}

\begin{figure}[t!]
\resizebox{0.95\hsize}{!}{\includegraphics{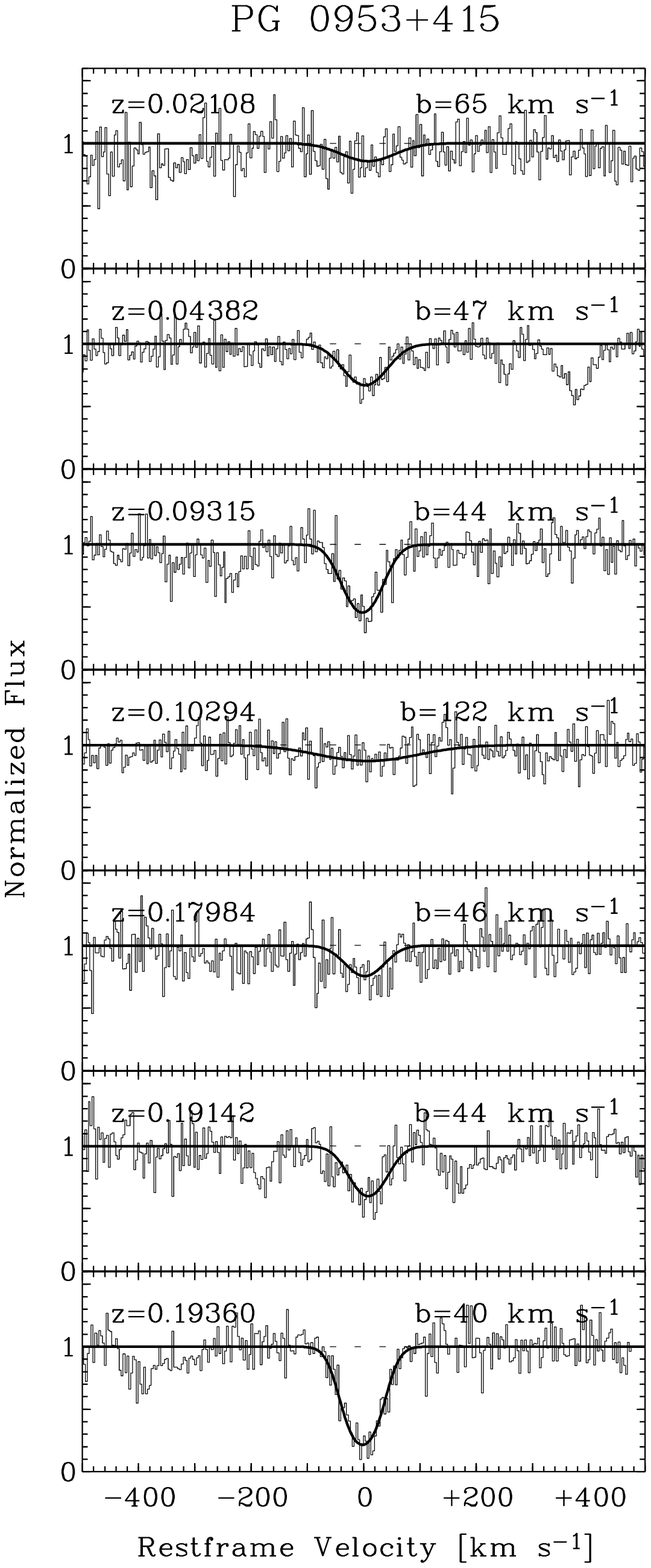}}
\caption[]{Broad Ly\,$\alpha$ absorbers towards PG\,0953+415, plotted
on a restframe velocity scale.
Voigt-profiles fits are indicated with the solid line. Other absorption
features are related to the Ly\,$\alpha$ forest and metal lines.}
\end{figure}

The wavelength range available for detecting intervening broad
Ly\,$\alpha$ absorption towards H\,1821+643 ($z_{\rm em}
=0.297$) is from $\sim 1230.0$ \AA\, to $\sim 1556.5$ \AA\,
($\geq 5000$ km\,s$^{-1}$ away from the quasar), corresponding
to a redshift interval of $\Delta z=0.269$. The typical S/N 
in this range is $\sim 21$ per resolution element, so that
the detection limit for broad Ly\,$\alpha$ absorption 
(see equation 4) is log $(N/b)\approx 11.2$. The blocking of 
certain spectral regions due to other interstellar and intergalactic 
lines reduces the total unblended redshift interval to a value of 
$\Delta z=0.238$. We find six
reliably detected single-component BLA candidates at 
$z=0.02642, 0.12221, 0.16352, 0.18047, 0.22616$, and
$0.26659$. Equivalent widths, column
densities, and $b$ values for these absorbers are listed 
in Table 2.
The systems at $z=0.16352$, $0.18047$, and $0.22616$
have been previously identified 
as H\,{\sc i} Ly\,$\alpha$ absorption in
lower-resolution
GHRS G140L data (Tripp, Lu, \& Savage 1998).
Two of these systems (at $z=0.22616$ and $0.26659$)
show associated O\,{\sc vi} absorption, which has been
previously analyzed by Tripp, Savage \&
Jenkins (2000). For the $z=0.26659$ system we find
associated Ly\,$\beta$ absorption, which is 
also broad ($b\approx42$ km\,s$^{-1}$).
Considering all six BLA candidates 
we obtain a BLA number density per unit redshift of 
$dN_{\rm BL}/dz \approx 25$ for log $(N/b)\geq 11.2$. 
As an example for broad Ly\,$\alpha$ absorption 
towards H\,1821+643 we show in Fig.\,3 the spectral region between
$1430$ and $1440$ \AA. The BLA
near $1435$ \AA\,($z=0.18047$) is clearly
visible, and there is no evidence that this feature
is produced by a continuum undulation or a blend
of several narrow absorption lines.
Continuum normalized velocity profiles of all 
BLA candidates towards H\,1821+643
are shown in Fig.\,4.
From Voigt profile fitting of the Ly\,$\alpha$ 
profiles we derive $b$ values between $44$ and $56$ km\,s$^{-1}$
and logarithmic H\,{\sc i} column densities between $13.19$ and
$13.65$ (see Table 2 for details). All the BLAs candidate
lines toward H\,1821+643 therefore are relatively narrow, which
may imply that some of them
are not related to the WHIM,
but rather are Ly\,$\alpha$ forest lines broadened by
non-thermal processes.
Our measured values of $b=45.6\pm 3.9$ km\,s$^{-1}$ and log $N=13.65\pm 0.06$
for the BLA at $z=0.26659$ agree well
with the numbers cited by Tripp, Savage, \& Jenkins (2000)
($b=44.6^{+7.3}_{-6.3}$ km\,s$^{-1}$ and log $N=13.62\pm 0.03$)
for this system. 
Note that we do not consider here the broad 
Ly\,$\alpha$ absorption found in the $z=0.12120$
system, as this line is part of a complex multiphase
absorption system with very broad O\,{\sc vi} absorption,
possibly indicating unresolved component structure and
non-thermal broadening processes (Tripp et al.\,2001).
A complete IGM analysis of the H\,1821+643 sight line will be 
provided in a separate paper (Sembach et al.\,2005a).

\subsection{PG\,0953+415}

Towards PG\,0953+415 we detect seven BLA candidates
at $z=0.02108$, $0.04382$, $0.09315$, $0.10294$, 
$0.17984$, $0.19142$, and $0.19360$
in the available wavelength range between $\sim 1220.0$ and $\sim
1485.9$ \AA\, ($\Delta z=0.219$).  
After applying a blocking
correction, the total redshift path for finding BLAs
towards PG\,0953+415 reduces to $\Delta z=0.202$.
The systems at $z=0.02108$, $0.10294$, and $0.19142$ are
tentative detections (see Table 2, last column for a more detailed
description of these lines). Another broad spectral
feature is observed near $1282$ \AA, corresponding 
to $z=0.05461$ for H\,{\sc i} Ly\,$\alpha$.
This feature most likely is an order-edge artifact caused by
the slightly imperfect ripple correction in the STIS data
(see Sect.\,2.4) and will not be considered any further.
The S/N in the PG\,0953+415 spectrum is $\sim 12$, implying
a typical detection limit for broad Ly\,$\alpha$ absorption
of log $(N/b)\approx 11.4$ (equation 4).
The systems at 
$z=0.09315$, $0.17984$, $0.19142$, and $0.19360$
have been previously 
identified as H\,{\sc i} Ly\,$\alpha$ by Savage et al.\,(2002).
None of the broad Ly\,$\alpha$ candidates towards PG\,0953+415
shows associated O\,{\sc vi} absorption; the two O\,{\sc vi}
absorbers identified by Savage et al.\,(2002) are both multiphase
absorbers with H\,{\sc i} Lyman series absorption having $b<40$ km\,s$^{-1}$.
The system at $z=0.19360$ is also detected in broad Ly\,$\beta$
absorption (with a line width consistent with that of the 
Ly\,$\alpha$ absorption).
From the four reliably detected BLA candidates 
and the blocking-corrected redshift path of $\Delta z=0.202$
we obtain $dN_{\rm BL}/dz \approx 20$ 
for log $(N/b)\geq 11.3$ for the line of sight towards
PG\,0953+415. Including the three tentative cases
we find $dN_{\rm BL}/dz \approx 35$.
All seven systems are shown in Fig.\,5.
Voigt profile fitting of these lines results in
$b$ values between $40$ and $122$ km\,s$^{-1}$ and logarithmic
H\,{\sc i} column densities between $12.88$ and $13.91$, as
listed in Table 2.

\begin{table*}[]
\caption[]{BLA candidates towards H\,1821+643, PG\,0953+415, PG\,1116+215, and PG\,1259+593}
\begin{tabular}{ccrrrcll}
\hline
$\lambda_{\rm obs}$ & $z$ &  $W_{r}$\,$^{\rm a}$ & log $N$\,$^{\rm a}$ & 
$b$\,$^{\rm a}$ & log $(N/b)$ & Other det. & Status\,$^{\rm b}$\\
$[$\AA$]$       &   & $[$m\AA$]$    &    & [km\,s$^{-1}$]  & & species \\
\hline

\multicolumn{8}{c}{H\,1821+643}\\
\hline
1247.789 &    0.02642 &  $79\pm10$ & $13.24\pm0.07$ & $51.3\pm6.0$    &  11.5 &  & \\
1364.232 &    0.12221 &  $74\pm8$  & $13.23\pm0.05$ & $44.4\pm3.8$    &  11.6 &  & \\ 	
1414.461 &    0.16352 &  $75\pm9$  & $13.19\pm0.06$ & $55.7\pm4.9$    &  11.4 &  & \\
1435.057 &    0.18047 &  $68\pm9$  & $13.21\pm0.08$ & $52.8\pm5.4$    &  11.5 &  & \\
1490.605 &    0.22616 & $139\pm14$ & $13.47\pm0.07$ & $53.8\pm5.7$    &  11.7 & (O\,{\sc vi}) & \\
1539.750 &    0.26659 & $174\pm13$ & $13.65\pm0.06$ & $45.6\pm3.9$    &  12.0 & Ly\,$\beta$, O\,{\sc vi} & \\
\hline
\multicolumn{8}{c}{PG\,0953+415}\\
\hline
1241.302 &    0.02108 &  $69\pm11$ & $13.12\pm0.10$ & $65.1\pm9.1$    &  11.3 &  & tentative, S\\
1268.938 &    0.04382 & $119\pm15$ & $13.34\pm0.07$ & $47.2\pm4.2$    &  11.7 &  & \\
1328.905 &    0.09315 & $194\pm16$ & $13.66\pm0.06$ & $43.9\pm5.3$    &  12.0 &  & \\ 
1340.807 &    0.10294 & $114\pm18$ & $13.32\pm0.16$ & $121.7\pm22.4$  &  11.2 &  & tentative, S\\ 
1434.275 &    0.17982 &  $74\pm10$ & $13.24\pm0.07$ & $46.0\pm5.5$    &  11.6 &  & \\ 
1448.378 &    0.19142 & $126\pm10$ & $13.47\pm0.08$ & $43.8\pm3.7$    &  11.8 &  & tentative, A\\
1451.026 &    0.19360 & $232\pm10$ & $13.91\pm0.04$ & $40.3\pm2.8$    &  12.3 & Ly\,$\beta$ & \\ 
\hline
\multicolumn{8}{c}{PG\,1116+215}\\
\hline
1235.550 &    0.01635 &  $111\pm10$ & $13.39\pm0.06$ & $48.5\pm5.1$   &  11.7 &  &  \\
1265.820 &    0.04125 &  $78\pm16$  & $13.24^{+0.88}_{-0.11}$ & $105.0\pm18.0$   & 11.2  & O\,{\sc vi} & tentative, M\\ 
1289.490 &    0.06072 &  $80\pm8$   & $13.28\pm0.06$ & $55.4\pm5.8$   &  11.5 &  & \\
1291.580 &    0.06244 &  $74\pm9$   & $13.18\pm0.07$ & $77.3\pm9.0$   &  11.3 & O\,{\sc vi} & \\
1320.060 &    0.08587 &  $18\pm7$   & $12.90\pm0.10$ & $52.0\pm14.0$  &  11.2 &  & tentative, S\\
1328.470 &    0.09279 &  $111\pm14$ & $13.39\pm0.09$ & $133.0\pm17.0$ &  11.3 &  & tentative, B\\
1378.210 &    0.13370 &  $86\pm11$  & $13.27\pm0.08$ & $83.6\pm10.4$  &  11.3 &  & tentative, A\\
\hline
\multicolumn{8}{c}{PG\,1259+593}\\
\hline
1218.449 &    0.00229 &  $190\pm24$ & $13.57\pm0.10$ & $42.1\pm4.4$   &  11.9 & Ly\,$\beta$, O\,{\sc vi} &  \\
1221.024 &    0.00440 &  $215\pm24$ & $13.65\pm0.14$ & $142.6\pm29.4$ &  11.5 &  & tentative, S\\
1221.812 &    0.00505 &  $133\pm20$ & $13.44\pm0.10$ & $57.2\pm8.9$   &  11.7 &  & tentative, S\\
1232.046 &    0.01347 &   $90\pm17$ & $13.24\pm0.14$ & $104.1\pm17.5$ &  11.2 &  & tentative, S\\
1267.334 &    0.04250 &   $58\pm7$  & $13.06\pm0.06$ & $40.7\pm4.9$   &  11.5 &  & tentative, A\\
1283.572 &    0.05586 &  $149\pm20$ & $13.46\pm0.14$ & $144.2\pm27.9$ &  11.3 &  & tentative, A\\
1298.456 &    0.06810 &   $72\pm10$ & $13.14\pm0.09$ & $87.7\pm9.2$   &  11.2 &  & tentative, M\\
1313.423 &    0.08041 &   $45\pm7$  & $12.97\pm0.10$ & $42.0\pm4.5$   &  11.3 &  & \\
1327.468 &    0.09196 &   $72\pm22$ & $13.13\pm0.21$ & $113.4\pm29.2$ &  11.1 &  & tentative, A\\
1340.654 &    0.10281 &  $136\pm18$ & $13.41\pm0.17$ & $196.9\pm31.2$ &  11.1 &  & tentative, A\\
1377.977 &    0.13351 &   $53\pm11$ & $13.01\pm0.08$ & $48.1\pm4.3$   &  11.3 &  & tentative, A\\
1386.276 &    0.14034 &   $60\pm10$ & $13.06\pm0.07$ & $56.7\pm5.4$   &  11.3 &  & tentative, S\\
1390.498 &    0.14381 &   $80\pm15$ & $13.18\pm0.13$ & $115.8\pm9.6$  &  11.1 &  & tentative, S\\
1396.225 &    0.14852 &  $254\pm9$  & $13.91\pm0.06$ & $42.1\pm2.4$   &  12.3 & Ly\,$\beta$ & \\
1399.668 &    0.15136 &  $106\pm9$  & $13.32\pm0.09$ & $65.3\pm5.5$   &  11.5 &  & \\
1418.044 &    0.16647 &  $113\pm13$ & $13.34\pm0.08$ & $93.3\pm8.9$   &  11.4 &  & tentative, A, R\\
1424.130 &    0.17148 &   $94\pm19$ & $13.25\pm0.16$ & $131.5\pm17.4$ &  11.1 &  & tentative, M, R\\
1433.167 &    0.17891 &  $101\pm10$ & $13.29\pm0.10$ & $98.5\pm9.1$   &  11.3 &  & tentative, M\\
1441.174 &    0.18550 &   $77\pm11$ & $13.17\pm0.12$ & $86.4\pm10.0$  &  11.2 &  & tentative, A, R\\
1493.587 &    0.22861 &  $133\pm10$ & $13.47\pm0.05$ & $40.3\pm2.9$   &  11.9 &  & \\
1508.963 &    0.24126 &  $130\pm12$ & $13.41\pm0.09$ & $89.1\pm6.9$   &  11.5 &  & \\
1566.428 &    0.28853 &  $111\pm16$ & $13.34\pm0.11$ & $70.7\pm8.5$   &  11.5 &  & tentative, S\\
1591.389 &    0.30906 &   $78\pm9$  & $13.20\pm0.10$ & $45.3\pm5.0$   &  11.5 &  & tentative, S\\
1600.816 &    0.31682 &   $62\pm14$ & $13.05\pm0.17$ & $43.5\pm6.4$   &  11.4 &  & tentative, S\\
1604.410 &    0.31978 &  $389\pm16$ & $13.99\pm0.09$ & $74.4\pm8.7$   &  12.1 & Ly\,$\beta$, O\,{\sc vi} & \\
1610.496 &    0.32478 &   $86\pm12$ & $13.24\pm0.15$ & $46.1\pm10.2$  &  11.6 &  & tentative, M\\
1676.519 &    0.37909 &  $116\pm19$ & $13.36\pm0.15$ & $72.0\pm8.9$   &  11.5 &  & tentative, S\\
1680.862 &    0.38266 &  $403\pm77$ & $13.92\pm0.42$ & $200.1\pm22.8$ &  11.6 &  & tentative, M\\
1723.654 &    0.41786 &   $88\pm10$ & $13.25\pm0.08$ & $50.7\pm3.9$   &  11.5 &  & tentative, A\\
\hline
\end{tabular}

$^{\rm a}$ $1\sigma$ errors are given.\\
$^{\rm b}$ Lines listed as tentative are all clearly detected with 
$W_{\rm r}/ \sigma >4.4$ except for the $z=0.08587$ line toward PG\,1116+215
with $W_{\rm r}/ \sigma =2.6$. However, the tentative lines may not have the required Gaussian
line shape to be considered as good candidates for single component thermally broadened BLAs.
The following abbreviations are used to explain why the detected lines are marked tentative: 
A = line is possibly asymmetric; B = line is blended; 
M = evidence for multi-component substructure; R = possible artifact caused by the MAMA repeller wire; 
S = S/N in the profile too low to reliably evaluate possible component structure.

\end{table*}

\subsection{PG\,1116+215 and PG\,1259+593}

Detailed analyses of the BLAs
towards PG\,1116+215 and PG\,1259+593 have been presented
by Sembach et al.\,(2004) and Richter et al.\,(2004), respectively. We 
here briefly summarize and update the findings from these studies. 
Toward PG\,1116+215, Sembach et al.\,(2004) find seven BLA
candidates along a total,
unblocked redshift path of $\Delta z=0.133$. 
We regard four of these absorbers as tentative detections
(see Table 2, last column).
With an 
average S/N of $\sim 17$, the study implies 
$dN_{\rm BL}/dz \approx 23$
for log $(N/b)\geq 11.2$ for the reliably detected systems 
($\sim 53$ including the uncertain cases).
Toward PG\,1259+593 there are 29 BLA candidates
along relatively long, unblocked redshift path of 
$\Delta z=0.355$ (Richter et al.\,2004). Following
the selection criteria presented above, 22 of these
candidate BLA systems are tentative detections.
The average S/N in the 
the PG\,1259+593 spectrum is $12$, but
note that the S/N varies within a factor of $\sim 1.5$ along the
spectrum.  
The number of reliably detected broad systems ($=7$)
corresponds to $dN_{\rm BL}/dz \approx 20$ for 
log $(N/b)\geq 11.1-11.4$ ($\sim 82$ including the 
uncertain cases). Column densities, $b$ values
and other information about these broad Ly\,$\alpha$
candidates towards PG\,1116+215 and PG\,1259+593 are
listed in Tables 2 and 3. 

\section{Properties of low-$z$ broad Ly\,$\alpha$ absorbers}

\subsection{Number distribution}

The measured values for $dN_{\rm BLA}/dz$ along each individual
line of sight toward H\,1821+643, PG\,0953+415, PG\,1116+215, and
PG\,1259+593 are summarized in Table 5. For the definitely detected
systems, $dN_{\rm BLA}/dz$ varies between 20 (PG\,0953+415 and PG\,1259+593) and
25 (H\,1821+643). If we include the tentative detections, $dN_{\rm BLA}/dz$
ranges between 25 (H\,1821+643) and 82 (PG\,1259+593). 
The discrepancy between the values for $dN_{\rm BLA}/dz$ 
along the individual sight lines may be due to intrinsic scatter 
from the low-number statistics, possibly reflecting the peculiarities 
of the WHIM distribution along individual sight lines. On the other
hand, part of the scatter may be caused by the varying sensitivity
to detect broad Ly\,$\alpha$ absorption in these spectra due to
the varying S/N. The PG\,1259+593 sight line, for instance,
provides a large number of tentative BLA detections in regions
with relatively low S/N. Most likely, a fair fraction of these
tentative BLA candidates are not WHIM absorbers but are
caused by noise and unresolved component structure.
Combining these four lines of sight, we have a total redshift 
path of $\Delta z=0.928$ to detected BLA candidates.
The comoving total path length $\Delta X$ 
sums up to $\Delta X=1.071$, as calculated
from equation (6). Along this redshift path, the STIS data then yield
a mean value of $dN_{\rm BLA}/dz=22\,(53)$ 
\footnote{22 for the secure detections, 53 for entire candidate sample
including the tentative cases.} for absorbers with log $(N/b)>11.1$ to $11.4$,
with an absorber weighted average of log $(N/b)\gtrsim 11.3$.

\subsection{Identification of different absorber types}

We can distinguish between
BLAs that are detected solely in Ly\,$\alpha$ and 
systems that show associated absorption in Ly\,$\beta$ and/or O\,{\sc vi}.
Associated Ly\,$\beta$ and/or O\,{\sc vi} absorption
may be of great importance to gain
insight into the physical conditions of these systems. In our sample
there are eight BLA systems that show associated absorption 
by Ly\,$\beta$ and/or O\,{\sc vi}, as indicated in Table 2, seventh column.
In Table 3 we summarize detailed information about these systems, such
as H\,{\sc i} and O\,{\sc vi} $b$ values and
O\,{\sc vi} velocity offsets.

\begin{table*}[t!]
\caption[]{BLA systems with associated Ly\,$\beta$/O\,{\sc vi} absorption}
\begin{tabular}{rllrrrrr}
\hline
$z$ & Sight line & Det.\,lines & $b_{\rm \,Ly\,\alpha}$ & 
$b_{\rm \,Ly\,\beta}$\,$^{\rm a}$ & $b_{\rm \,O\,VI}$\,$^{\rm a}$
& $\Delta v_{\rm \,O\,VI}$ & $b_{\rm \,H\,I}/b_{\rm \,O\,VI}$ \\ 
&      &      & [km\,s$^{-1}$]        & [km\,s$^{-1}$]
& [km\,s$^{-1}$]  & [km\,s$^{-1}$] & \\
\hline

0.22616 & H\,1821+643 &    H\,{\sc i}\,$\lambda 1215.7$, O\,{\sc vi}\,$\lambda\lambda 1031.9, 1037.6$ &
          50.8 & ... & 15.7 & 60 & 3.2 \\
0.26659 & H\,1821+643 &    H\,{\sc i}\,$\lambda\lambda 1215.7, 1025.7$, O\,{\sc vi}\,$\lambda\lambda 1031.9, 1037.6$ &
          45.6 & 42.2 & 24.1 & 9 & 1.9 \\
0.19360 & PG\,0953+415 &   H\,{\sc i}\,$\lambda\lambda 1215.7, 1025.7$ &
          40.3 & ... & ... & ... & ... \\
0.04125 & PG\,1116+215 &   H\,{\sc i}\,$\lambda 1215.7$, O\,{\sc vi}\,$\lambda\lambda 1031.9, 1037.6$ &
          105.0 & ... & ... & 35 & ... \\
0.06244 & PG\,1116+215 &   H\,{\sc i}\,$\lambda 1215.7$, O\,{\sc vi}\,$\lambda\lambda 1031.9, 1037.6$ &
          77.3 & ... & ... & 20 & ... \\
0.00229 & PG\,1259+593 &   H\,{\sc i}\,$\lambda\lambda 1215.7, 1025.7$, O\,{\sc vi}\,$\lambda\lambda 1031.9$ &
          42.1 & ... & ... & 48 & ... \\
0.14852 & PG\,1259+593 &   H\,{\sc i}\,$\lambda\lambda 1215.7, 1025.7$ &
          42.1 & ... & ... & ... & ... \\
0.31978 & PG\,1259+593 &   H\,{\sc i}\,$\lambda\lambda 1215.7, 1025.7$, O\,{\sc vi}\,$\lambda\lambda 1031.9, 1037.6$ &
          74.3 & 62.7 & 19.3 & 10 & 3.9 \\
\hline
\end{tabular}
\noindent
$^{\rm a}$ $b$ values measured only for lines in the STIS wavelength regime.
\end{table*}

\subsubsection{BLAs with associated Ly\,$\beta$ absorption}

Five broad Ly\,$\alpha$ systems in our sample have an associated
broad Ly\,$\beta$ line. For the occurrence of associated
Ly\,$\beta$ absorption a minimum strength of log $(N/b)\approx 11.8$
is required. This means that the high-column density systems 
can be detected in Ly\,$\beta$, if this line falls into a spectral
region not blended by other ISM/IGM lines. For a thermally broadened WHIM
absorber, the $b$ value measured for Ly\,$\beta$ should be consistent
with that of the Ly\,$\alpha$ absorption.
If the $b$ values for Ly\,$\alpha$ and Ly\,$\beta$ are largely inconsistent 
with each other, blending effects or noise features are most likely
affecting the data and these systems should not be considered any
further. An example for this is the $z=0.30434$ system toward
PG\,1259+593 (see Richter et al.\,2004, their Fig.\,3f), which shows
a relatively broad Ly\,$\alpha$ line with $b\approx 65$ km\,s$^{-1}$, whereas
the Ly\,$\beta$ absorption is narrow with $b\approx 35$ km\,s$^{-1}$. For this
system we can conclude that blending and noise effects but not
thermal line broadening determine the observed large width 
of the Ly\,$\alpha$ absorption. This system therefore is not
included in our sample. A counter-example
is the BLA at $z=0.31978$ along the same line of sight,
which has a $b$ value of $\sim 70 $ km\,s$^{-1}$ consistent with both
Ly\,$\alpha$ and Ly\,$\beta$ absorption (Richter et al.\,2004, their Fig.\,3g).
Along all four sight lines we find nine BLAs
with $b\geq 40$ km\,s$^{-1}$ that 
have associated Ly\,$\beta$ absorption, but only for five of these systems is
the width of the corresponding Ly\,$\beta$ line consistent with 
$b\geq 40$  km\,s$^{-1}$. Although our statistics are poor, this 
implies that S/N features, blending issues, and other processes
may contaminate our BLA sample up to a level of
$\sim 50$ percent. This value is consistent with our
estimate from Sect.\,2.4, where we analyzed the influence
of noise in a set of artificial spectra.
	
\subsubsection{BLAs with associated O\,{\sc vi} absorption}

From the 20 reliably detected BLAs towards
H\,1821+643, PG\,0953+415, PG\,1116+215, and PG\,1259+593,
six show associated O\,{\sc vi} absorption within 
a velocity range of $\pm 60$ km\,s$^{-1}$. 
Only a small fraction of the broad Ly\,$\alpha$
systems is expected to have associated O\,{\sc vi} absorption,
as the occurrence of O\,{\sc vi} critically depends
on the metallicity of the absorbing gas, its temperature,
and its total column density. Therefore, most
BLAs arise in WHIM gas that is
too hot, to metal-poor, or too diffuse for O\,{\sc vi} 
absorption to occur at detectable levels.
Some of these ``hybrid systems'' that have good S/N in both
Ly\,$\alpha$ and O\,{\sc vi} absorption have been analyzed
in detail with respect to their physical properties, such
as temperature, ionization conditions, and metallicity
(Sembach et al.\,2004; Richter et al.\,2004; Tripp, Savage,
\& Jenkins 2000). Under the assumption of collisional
ionization equilibrium, the column densities of H\,{\sc i}
and O\,{\sc vi} can be used to constrain the oxygen abundance
in the gas (see, e.g., the $z=0.31978$ system towards
PG\,1259+593; Richter et al.\,2004). The measured
$b$ values for H\,{\sc i} and O\,{\sc vi} also can provide information
about the influence of non-thermal broadening components in 
some systems. The comparison between broad H\,{\sc i} and
O\,{\sc vi} absorption is often difficult, however, as the 
observed O\,{\sc vi} features do not exactly align in
velocity space with the broad Ly\,$\alpha$ absorption (see
also Sect.\,2.4).
The velocity centroids of
broad Ly\,$\alpha$ and O\,{\sc vi} in these six systems
deviate by $9-60$ km\,s$^{-1}$ 
(see Table 3). Such velocity offsets could be caused
by substructures within the filament.
It is plausible that WHIM filaments are neither perfectly homogeneous
(in terms of volume densities), nor are they perfectly isothermal
(Kawahara et al.\,2005). If 
substructures with higher densities exist, the gas may be able to
cool, creating distinct regions where O\,{\sc vi}
absorption may occur preferentially once the gas reaches the
peak temperature of O\,{\sc vi} near $3\times10^5$ K. The
radial velocities of these denser spots would most likely
differ from the center velocity of the entire filament, in
particular if the gas stays
in a hydrostatic equilibrium (e.g., Valageas, Schaeffer, \& Silk 2002). Cooler spots 
then would move towards the mass center of the filament,
creating a cooling flow. Such effects may lead to
a non-uniform optical depth distribution of H\,{\sc i} and O\,{\sc vi}
along the WHIM filament
and could produce velocity offsets between the integrated 
absorption profiles of these two species.
It therefore remains unclear whether the observed broad H\,{\sc i} and the
O\,{\sc vi} trace the same gas phase. 

Also the comparison between H\,{\sc i} and O\,{\sc vi} $b$ values
provides information about possible non-thermal broadening effects,
as outlined in Sect.\,2.3. For three out of the six H\,{\sc i}/O\,{\sc vi}
absorber pairs accurate O\,{\sc vi} $b$ values can be measured and the 
resulting $b_{\rm H\,I}/b_{\rm O\,VI}$ ratios are listed in the last
row of Table 3. Only the system at $z=0.31978$ towards PG\,1259+593
has $b_{\rm H\,I}\approx 4\,b_{\rm O\,VI}$, as expected for pure
thermal line broadening (equation 2).
The systems at $z=0.22616$ and $z=0.26659$ towards H\,1821+643
both have $b_{\rm H\,I}< 4\,b_{\rm O\,VI}$, thus indicating
that $b_{\rm non-th}$ dominates over $b_{\rm th}$ in these systems.
The large widths of the
O\,{\sc vi} lines could be caused by unresolved substructure, while
the multi-component O\,{\sc vi} still could arise in hot gas that produces 
thermally broadened H\,{\sc i} absorption.
Since it is unclear whether the broad Ly\,$\alpha$ 
absorption and the O\,{\sc vi} arise in the same gas phase, we
have to conclude that the 
comparison between BLA and O\,{\sc vi} line widths
remains inconclusive.
Two of the six broad Ly\,$\alpha$/O\,{\sc vi} systems 
occur in the line of sight towards H\,1821+643.
This sight line exhibits
a overabundance of intervening O\,{\sc vi} systems
compared to other low-z lines of sight (see
Savage et al.\,2002; Tripp, Savage, \& Jenkins 2000).
However, we here find that the H\,1821+643 sight line
is not particularly rich in broad Ly\,$\alpha$
candidate systems, when compared to 
PG\,0953+415, PG\,1116+215, and PG\,1259+593. 

\begin{table}
\caption[]{CIE properties of broad Ly\,$\alpha$ absorbers}
\begin{tabular}{ccccc}
\hline
$z$ & log $T$ & log $f_{\rm H}$ & log $N$(H)\,$^{\rm a}$ & Status\,$^{\rm b}$ \\
\hline
\multicolumn{5}{c}{H\,1821+643}\\
\hline
0.02642 & 5.20 & 5.25 & 18.49 & \\
0.12221 & 5.07 & 5.00 & 18.23 & \\
0.16352 & 5.27 & 5.39 & 18.58 & \\
0.18047 & 5.22 & 5.30 & 18.51 & \\
0.22616 & 5.24 & 5.33 & 18.80 & \\
0.26659 & 5.10 & 5.05 & 18.70 & \\
\hline
\multicolumn{5}{c}{PG\,0953+415}\\
\hline
0.02108 & 5.41 & 5.65 & 18.77 & tentative\\
0.04382 & 5.13 & 5.11 & 18.45 & \\
0.09315 & 5.06 & 4.98 & 18.64 & \\
0.10294 & 5.95 & 6.55 & 19.87 & tentative\\
0.17982 & 5.10 & 5.06 & 18.30 & \\
0.19142 & 5.06 & 4.98 & 18.45 & tentative\\
0.19360 & 4.99 & 4.83 & 18.74 & \\
\hline
\multicolumn{5}{c}{PG\,1116+215}\\
\hline
0.01635 & 5.15 & 5.16 & 18.55 & \\
0.04125 & 5.82 & 6.35 & 19.59 & tentative\\
0.06072 & 5.27 & 5.38 & 18.66 & \\
0.06244 & 5.55 & 5.91 & 19.09 & \\
0.08587 & 5.21 & 5.28 & 18.18 & tentative\\
0.09279 & 6.03 & 6.66 & 20.05 & tentative\\
0.13370 & 5.62 & 6.03 & 19.30 & tentative\\
\hline
\multicolumn{5}{c}{PG\,1259+593}\\
\hline
0.00229 & 5.03 & 4.91 & 18.48 & \\
0.00440 & 6.09 & 6.74 & 20.39 & tentative\\
0.00505 & 5.29 & 5.44 & 18.88 & tentative\\
0.01347 & 5.81 & 6.34 & 19.58 & tentative\\
0.04250 & 5.00 & 4.84 & 17.90 & tentative\\
0.05586 & 6.10 & 6.76 & 20.22 & tentative\\
0.06810 & 5.66 & 6.10 & 19.24 & tentative\\
0.08041 & 5.02 & 4.90 & 17.87 & \\
0.09196 & 5.89 & 6.45 & 19.58 & tentative\\
0.10281 & 6.37 & 7.10 & 20.51 & tentative\\
0.13351 & 5.14 & 5.14 & 18.15 & tentative\\
0.14034 & 5.29 & 5.42 & 18.48 & tentative\\
0.14381 & 5.91 & 6.48 & 19.66 & tentative\\
0.14852 & 5.03 & 4.91 & 18.82 & \\
0.15136 & 5.41 & 5.65 & 18.97 & \\
0.16647 & 5.72 & 6.19 & 19.53 & tentative\\
0.17148 & 6.02 & 6.64 & 19.89 & tentative\\
0.17891 & 5.77 & 6.26 & 19.55 & tentative\\
0.18550 & 5.65 & 6.08 & 19.26 & tentative\\
0.22861 & 4.99 & 4.83 & 18.30 & \\
0.24126 & 5.68 & 6.12 & 19.53 & \\
0.28853 & 5.48 & 5.78 & 19.12 & tentative\\
0.30906 & 5.09 & 5.04 & 18.24 & tentative\\
0.31682 & 5.06 & 4.96 & 18.01 & tentative\\
0.31978 & 5.52 & 5.85 & 19.84 & \\
0.32478 & 5.11 & 5.07 & 18.31 & tentative\\
0.37909 & 5.49 & 5.80 & 19.16 & tentative\\ 
0.38266 & 6.38 & 7.12 & 21.04 & tentative\\ 
0.41786 & 5.19 & 5.23 & 18.48 & tentative\\
\hline

\end{tabular}

$^{\rm a}$ Total hydrogen column densities are derived under the assumption
that the gas is in collisional ionization equilibrium (i.e., additional photoionization
is not considered).\\
$^{\rm b}$ The tentative status implies that the BLA, although clearly detected, 
may not qualify as a single-component Gaussian broadened
absorber for the reasons given in Table 2.

\end{table}

\subsection{CIE properties of the BLAs}

Despite the remaining uncertainty about the interpretation
of the the observed line widths of BLAs and the uncertain contribution
of photoionization, we now want to estimate
the total gas content of these absorbers assuming the ideal case in which
the lines are predominantly thermally broadened while the gas 
is in collisional ionization equilibrium. We 
use the measured $b$ values and H\,{\sc i} column densities 
from our STIS sample together with equations (2) and (3) to derive the total
hydrogen column density, $N$(H), in these systems. In Table 4 we have
listed for each of the 49 BLA candidates values for log $T$, log $f_{\rm H}$, 
and log $N$(H). The BLAs cover a temperature range from $\sim 10^5$ to
$\sim 2.4 \times 10^6$ K for pure thermal broadening, and the resulting
total hydrogen column densities span a range from
$\sim 10^{18}$ to $\sim 10^{21}$ cm$^{-2}$. All of the
very broad systems with $b>100$ km\,s$^{-1}$ are considered to be
tentative detections (see Table 4), so that the validity of the highest column
density systems is particularly critical.

\begin{figure*}[t!]
\resizebox{1.0\hsize}{!}{\includegraphics{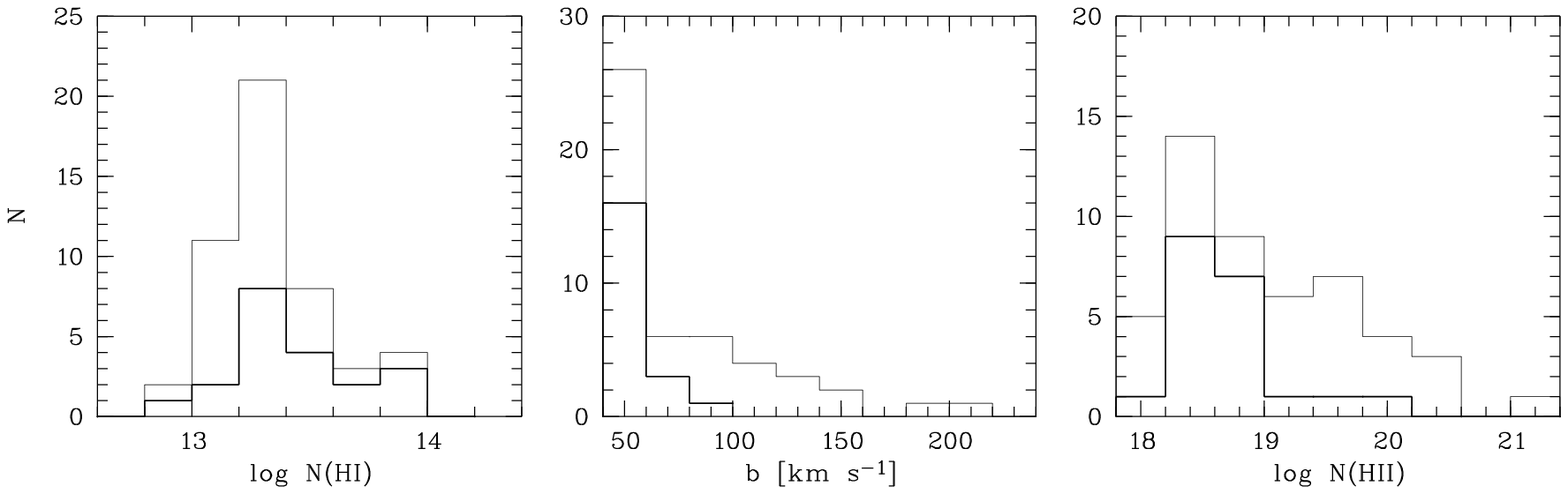}}
\caption[]{Number distributions of log $N$(H\,{\sc i}), $b$, and
log $N$(H\,{\sc ii}) for our BLA sample are shown. The thick line
indicates the distribution for the secure cases, the thin line shows
the distribution for the entire BLA sample.}
\end{figure*}

\subsection{Distribution of hydrogen column densities and $b$ values}

In Fig.\,6 we show the distribution of H\,{\sc i} column densities, $b$ values,
and (estimated) ionized hydrogen column densities ($N$(H\,{\sc ii}$)\approx N_{\rm total}$)
for our BLA sample. Thin lines indicate the distribution
for the entire candidate sample (49 absorbers), the thick lines indicate
the distribution for the secure cases (20 absorbers). The H\,{\sc i} column
density distribution (Fig.\,6, left panel) peaks in the range
log $N$(H\,{\sc i})$=13.2-13.4$. No system with
log $N$(H\,{\sc i})$\geq14.0$ has been found, demonstrating that the
neutral gas content of these absorbers is generally low. Note that
for log $N$(H\,{\sc i})$\leq13.0$ our sample is incomplete due to the
limited sensitivity in the lower S/N regions of the spectra.
Only a few systems are observed in this range, but it is probable
that the intrinsic broad Ly\,$\alpha$ number distribution rises further towards
the low-column density end.
The $b$ value distribution of the
BLAs (Fig.\,6, middle panel) is strongly peaked
at $b=40-60$ km\,s$^{-1}$ and shows a tail towards higher
$b$ values. The $b$ value distribution is difficult to
interpret, as there are various selection effects that have to be taken into
account. First of all, our sample is incomplete for higher $b$ values.
Since the sensitivity to detect broad Ly\,$\alpha$ lines scales with
$(N/b)$ (see equation 4), we preferentially detect narrow lines
rather than broad lines for a given column density level. 
It is probable that a considerable fraction of very broad lines lie
below our detection limit and do not
show up in the $b$ value distribution shown in Fig.\,6.
Secondly, if non-thermal line broadening mechanisms such as peculiar gas motions
are important also for the Ly\,$\alpha$ forest lines, we expect to
find a number of lines from the photoionized Ly\,$\alpha$ shifted
into the $b$ value regime between $40$ and $100$ km\,s$^{-1}$. Such
lines may significantly contaminate our broad Ly\,$\alpha$ sample at
low $b$ values, whereas the contamination should be less severe at
higher $b$ values. The number of WHIM absorbers
with $b$ values $b=40-60$ km\,s$^{-1}$ shown in Fig.\,6
thus may be overestimated.
At the sensitivity of the current data, we should be able to detect 
very broad ($b>100$ km\,$^{-1}$) Ly\,$\alpha$ absorption toward H\,1821+643, and 
possibly PG\,0953+415 down to a level of log $(N/b)\approx 11.2$ and $11.4$, respectively.  
However, all except one of the Ly\,$\alpha$ lines detected in     
these two directions have $b<100$ km\,$^{-1}$. This result stands in stark 
contrast to the PG\,1116+215 and PG\,1259+593 sight lines, for which 
several lines with $b>100$ km\,$^{-1}$ are found. Differences in 
the number of very broad Ly\,$\alpha$ absorbers for the four sight lines 
might be caused by geometrical effects resulting from the intersection
of the cosmic web filaments at different angles.  If so, this
would imply that the line broadening is dominated by flows within
the filaments or by projection effects. Alternatively, the four sight 
lines may intersect regions of the web with very different 
physical conditions, in which case the line-width differences may
be dominated by the temperature of the gas.  Distinguishing between
these two possibilities will require observations of a larger number
of sight lines to construct a complete line width frequency
distribution for absorbers in a wide variety of cosmic web environments.

In Fig.\,6, right panel, we show the distribution of
the ionized hydrogen column density, $N$(H\,{\sc ii}), as estimated
for each BLA candidate from equations (2) and (3) (see Sect.\,4.3 and Table 4).
If our estimate for $N$(H\,{\sc ii}) is at least roughly correct,
most of the detected WHIM BLA candidates have total gas column 
densities between $10^{18}-10^{20}$ cm$^{-2}$. Only three of the
securely detected systems have column densities larger than
$10^{19}$ cm$^{-2}$, while
for the more uncertain cases the relative fraction of systems
at these column densities is significantly higher.
This is because high $N$(H\,{\sc ii})
systems are expected to have the highest temperatures and thus
the lowest neutral gas content. As these systems will be
particularly broad and shallow, they are most difficult to detect.
Therefore, our sample is incomplete 
for total gas column densities $>10^{19}$ cm$^{-2}$.

\subsection{Broad Ly\,$\alpha$ absorbers and structures along the line of sight}

Important information about the origin of the broad Ly\,$\alpha$ absorption
can be obtained by comparing the redshift distribution of BLAs with that
of galactic structures along the individual lines of sight.
For the lines of sight toward H\,1821+643, PG\,0953+415, and PG\,1116+215
redshift information about nearby galaxies is available from 
previous studies (Tripp, Lu, and Savage 1998; Savage et al.\,2002).
These galaxy-redshift data have been obtained with the $3.5$m WIYN telescope
on Kitt Peak, using the multi-object fiber spectrograph (HYDRA). The 
measurements provide redshift information for galaxies in a field of
$\sim 1\deg ^2$ size around each quasar. Unfortunately, the completeness
of these galaxy redshift surveys varies substantially among the different
sight lines, mostly due to the varying stellar contamination. However, within
$\sim 30 \farcm 0$ of each quasar, the survey is $\sim 40$ percent complete
for objects with $B_J \leq20$ (see Tripp, Lu, and Savage 1998). 
In Fig.\,7 we compare the redshift
distribution of our BLA candidates to the distribution of galaxies along the
lines of sight towards H\,1821+643, PG\,0953+415, and PG\,1116+215 within
a radius of $30\farcm0$ from the QSO. From a visual inspection of Fig.\,7
it appears that BLAs are loosely correlated with the occurrence of 
galaxies along the redshift path. One or more galaxies are present 
within a velocity radius of 3000
km\,s$^{-1}$ around each BLA candidate.
However, in contrast to the O\,{\sc vi} absorbers (e.g., Sembach et al.\,2004)
the BLAs do not appear to be significantly clustered around the peaks of
the galaxy-redshift distribution. This suggests that the BLAs
also trace only mildly overdense WHIM filaments in the vicinity 
of individual galaxies, whereas the O\,{\sc vi} systems may 
be more closely related to higher density gas in
galaxy groups. 
However, we may have missed broad systems that are part of more
complex multi-component absorbers, as we have restricted our 
sample to single-component systems. 
Since multi-component systems most
likely arise preferentially in the vicinity of individual galaxies and
galaxy groups due to the more complex gas distribution in these environments,
our BLA sample is biased against WHIM absorbers close to galaxies 
and galaxy groups. Therefore, the observed weak correlation between BLAs
galaxies may simply be a result of our BLA selection criteria.
Additional galaxy-redshift data for other lines of sight 
(e.g., PG\,1259+593) will be required to improve the statistics and to
investigate the relation between BLAs and galaxies in more detail. 

\begin{figure}[t!]
\resizebox{0.9\hsize}{!}{\includegraphics{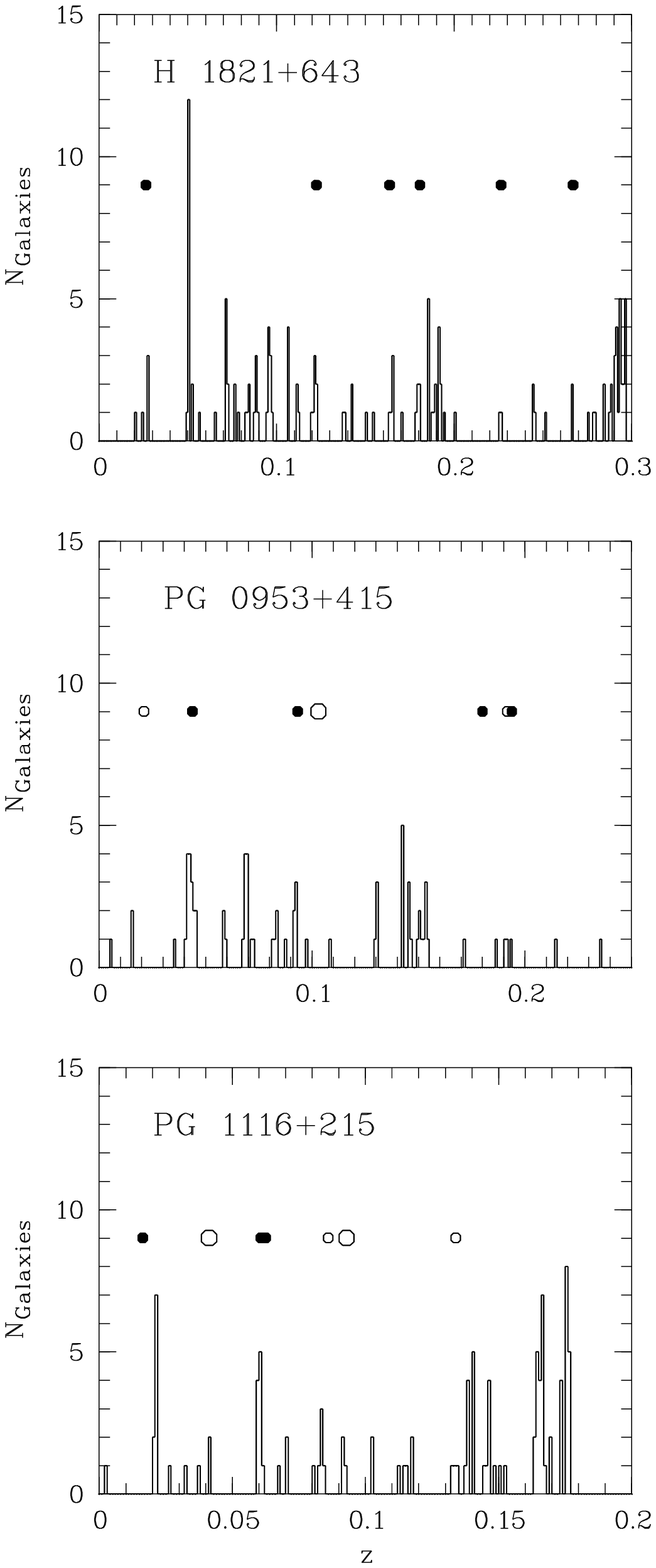}}
\caption[]{Distribution of galaxies within $\sim 30 \farcm 0$ of the direction
of H\,1821+643, PG\,0953+415, and PG\,1116+215. The bins have a
redshift width of $\Delta z=0.001$.
The redshifts of the BLAs
are indicated by filled circles (secure detections) and open
circles (tentative detections). The smaller circles indicate
BLA systems with $b\leq 100$ km\,s$^{-1}$, the larger circles
indicate very broad systems with $b> 100$ km\,s$^{-1}$.}
\end{figure}

\subsection{Estimate for $\Omega_b$}

Following equation (5), the total hydrogen column 
densities derived for each of the
detected BLA candidates toward H\,1821+643, PG\,0953+415, PG\,1116+215, and
PG\,1259+593 (Table 4) allow us to estimate the baryon
content of the broad Ly\,$\alpha$ absorbers. 
As discussed earlier in this paper, there is strong evidence that a significant fraction
of the detected BLA candidates have origins other than the WHIM, and that
the $b$-value distribution of the WHIM H\,{\sc i} absorbers is affected
also by non-thermal broadening processes. Moreover, we have shown
that photoionization from the UV background may significantly influence
the ionzation fraction in the BLAs.
However, as it is impossible to reliably quantify 
the contamination from these alternative processes, we now want
to provide an estimate for $\Omega_b$(WHIM) assuming the ideal case
(i.e., without any contamination and without photoionization). Summing over the hydrogen column
densities, log $N$(H), listed in Table 4, we obtain a total hydrogen column
density of log $N$(H)$=20.26$ for the reliably detected systems, 
and a value of log $N$(H)$=21.42$ including the tentative systems. 
The total path length to detect BLA along the four lines of sight is
$\Delta X=1.071$ (see Sect.\,4.1 and Table 5). From equation (6) we 
then derive $\Omega_b$(BLA)$=0.0027\,h_{70}\,^{-1}$ considering
only the 20 reliable cases 
and $\Omega_b$(BLA)$=0.0381\,h_{70}\,^{-1}$ for all 49 BLA candiates.
Estimates for $\Omega_b$(BLA) for each individual sight line are listed in
the last column of Table 5. 
Current cosmological models favour a value for $\Omega_b$ of $\sim0.045$,
and the contribution from the WHIM to $\Omega_b$ at $z=0$ is expected to
be $\sim 40$ percent (e.g., Dav\a'e et al.\,2001), 
so that $\Omega_b$(WHIM)$\approx 0.018$ is predicted in the local Universe. The value for
$\Omega_b$(BLA) derived from the restricted sample therefore is $\sim 16$ percent 
of the expected contribution from the WHIM to $\Omega_b$. This
is slightly higher than the contribution
from intervening O\,{\sc vi} absorbers 
($\Omega_b$(O\,{\sc vi})$\geq 0.0022$ $h_{70}\,^{-1}$; 
Danforth \& Shull 2005; Sembach et al.\,2004).
The estimate from the entire BLA candidate sample, however, exceeds the expected value for 
$\Omega_b$(WHIM) by a factor of $\sim 2$.
The reason for this is that the by far dominating contribution to $\Omega_b$(BLA) 
comes from the very broad lines, which are exclusively 
tentative detections. The reliability of these few very broad
features is very uncertain, and the unrealistically 
high value for $\Omega_b$(BLA) from these systems indeed implies
that a considerable fraction of these features probably are not
related to hot gas in the WHIM.
Note that broad lines with $b\geq100$ km\,s$^{-1}$ make 
up $\sim 90$ percent of the 
baryon content in the total sample, although 
they contribute only with $\sim25$ percent
to the total number of absorbers. 
If we take non-thermal broadening processes
and additional photoionization into account, we believe that our estimate of 
$\Omega_b$(BLA)$=0.0027\,h_{70}\,^{-1}$
from the limited data sample represents a realistic lower limit
for the baryon content of broad Ly\,$\alpha$ systems at low
$z$ for a detection threshold of log $(N/b)\approx 11.3$.

\section{Summary}

\begin{table*}[t!]
\caption[]{Sight line properties}
\begin{tabular}{lrrrrrlll}
\hline
Sight line & $z_{\rm em}$ & S/N\,$^{\rm a}$ & 
log $(N/b)_{\rm min}$\,$^{\rm b}$ &
$\Delta z_{\rm BLA}$ & $\Delta X_{\rm BLA}$ & $N_{\rm BLA}$ &
$dN_{\rm BLA}/dz$ & $\Omega_{b}$(BLA)\,$^{\rm c}$ \\
& & & & & & & & [$\times 10^{-2} h_{70}\,^{-1}$] \\
\hline
H\,1821+643  & $0.297$ & $21$ & $11.2$ & $0.238$ & $0.273$ & $6\,(6)\,^{\rm d}$ & $25\,(25)\,^{\rm d}$ & $0.13\,(0.13)\,^{\rm d}$ \\
PG\,0953+415 & $0.239$ & $12$ & $11.4$ & $0.202$ & $0.224$ & $4\,(7)$    & $20\,(35)$ & $0.08\,(0.55)$ \\
PG\,1116+215 & $0.177$ & $17$ & $11.2$ & $0.133$ & $0.144$ & $3\,(7)$    & $23\,(53)$ & $0.22\,(2.08)$ \\
PG\,1259+593 & $0.478$ & $12$ & $11.4$ & $0.355$ & $0.430$ & $7\,(29)$   & $20\,(82)$ & $0.45\,(8.37)$ \\
\hline
Total        &         &      &      & $0.928$ & $1.071$ & $20\,(49)$  & $22\,(53)$ & $0.27\,(3.81)$ \\
\hline
\end{tabular}
\noindent \\
\begin{scriptsize}
$^{\rm a}$ Typical S/N per resolution element ($\sim 7$ km\,s$^{-1}$).\\
$^{\rm b}$ Typical detection limit for BLAs. $N$ is in units cm$^{-2}$, $b$
in units km\,s$^{-1}$.\\
$^{\rm c}$ Estimates for $\Omega_{b}$(BLA) are calculated assuming pure thermal broadening for the BLAs and
collisional ionization equilibrium.\\
$^{\rm d}$ Plain numbers refer to the restricted BLA candidate sample, numbers put in parentheses refer to
the total BLA candidate sample (including the tentative cases).
\end{scriptsize}
\end{table*}

In this paper we have studied intervening broad
Ly\,$\alpha$ absorbers at low redshifts 
that probably trace large amounts of baryonic 
matter hidden in the
warm-hot intergalactic medium (WHIM) at temperatures
between $\sim 10^5$ and $\sim 10^6$ K.
In ionization 
equilibrium a very small fraction ($<10^{-5}$, 
typically) of the hydrogen in the WHIM is 
expected to be neutral. Depending
on parameters like the gas temperature and the total
gas column density, a certain fraction of this neutral gas should
be detectable as thermally broadened Ly\,$\alpha$
absorption at column densities log $N$(H\,{\sc i})$
\leq 14$ and $b\geq 40$ km\,s$^{-1}$.
Given the limitations in sensitivity for detecting
very broad and shallow absorption features in
currently existing STIS data, it becomes clear that
only certain regions of the WHIM (those with relatively low
temperatures and high total gas column densities) can be 
traced with broad Ly\,$\alpha$ absorption. If the
gas is in CIE and if thermal broadening dominates,
one can directly estimate the baryon content of
WHIM broad Ly\,$\alpha$ absorbers (BLAs) from the 
measured H\,{\sc i} line widths and column densities.
However, the CIE hypothesis probably is only a poor approximation
for low-density absorbers, so that the estimated ionization
fractions are uncertain by up to 50 percent.
In addition, the interpretation of BLA line widths is complicated, 
as non-thermal broadening processes like peculiar gas 
motions, macroscopic turbulence, the Hubble flow, as well
as noise features and continuum undulations potentially 
mimic some of the broad absorption features attributed
to the WHIM. 
Therefore, every WHIM filament that
contains sufficient amounts of neutral hydrogen will
produce broad Ly\,$\alpha$ absorption, but conversely, not every
broad spectral feature observed in QSO spectra is
related to the WHIM. Due to the non-thermal 
broadening processes, we expect that the error rate
for detecting the WHIM with broad Ly\,$\alpha$ absorption 
may be as high as 50 percent and
that for true WHIM BLAs the baryon content may be 
overestimated when relying solely on the observed $b$ value
distribution. However, it is clear that BLA systems represent an important
baryon reservoir in the local Universe. Thus, their
frequency in high-resolution QSO spectra and their 
physical properties should be studied to the greatest
extent possible.

Our analysis of the two STIS spectra of H\,1821+643 
and PG\,0953+415 has resulted in the detection 
of 14 broad Ly\,$\alpha$ absorber candidates
with $12.88\leq$\, log $N$(H\,{\sc i})$\leq 13.91$
and $40 \leq b \leq 122$ km\,s$^{-1}$ along a
total (unblocked) redshift path of $\Delta z = 0.440$.
These measurements complement previous analyses of 
broad Ly\,$\alpha$ absorption towards 
PG\,1259+593 (Richter et al.\,2004) and
PG\,1116+215 (Sembach et al.\,2004). Considering all four
lines of sight, we find 20 reliably detected 
BLA systems and additional 29 tentative cases 
along a total redshift path of $\Delta z=0.928$. 
With $20\,(49)$ detected BLA candidate systems, the
number of systems per unit redshift is $dN_{\rm BLA}/dz=
22\,(53)$. Most of these systems have H\,{\sc i} column densities
between log $N=13.0-13.6$ and $b$ values 
$40\leq b \leq 100$ km\,s$^{-1}$.
Assuming CIE and pure thermal line broadening, the total
hydrogen column density in these absorbers typically
ranges between log $N=18-20$. Unfortunately, most of the systems that
have relatively high total hydrogen column densities log $N\geq19$ (and thus 
the largest baryon content) belong to the tentative
detections, and thus are highly uncertain. 

Five of the stronger BLA systems show 
associated broad Ly\,$\beta$ absorption. The presence of
broad Ly\,$\beta$ is of great help to assure
that the H\,{\sc i} absorption in these systems 
is indeed {\it intrinsically} broad and not
caused by noise features and continuum undulations.
Six BLA candidate systems show associated O\,{\sc vi} absorption.
In all cases, the O\,{\sc vi} absorption shows
a velocity offset compared to the H\,{\sc i} absorption.
Moreover, in two out of three cases the measured 
O\,{\sc vi} $b$ values are
incompatible with the $b$ values measured for H\,{\sc i},
possibly implying unresolved component structure and
non-thermal line broadening. However, as it is not clear
whether or not the BLAs and the O\,{\sc vi} systems trace
the same gas phase, the comparison between these
two WHIM absorber types remains inconclusive.
If we consider only the 20 reliably detected BLA candidates,
their baryon content sums up to a value of 
$\Omega_b$(BLA)$=0.0027\,h_{70}\,^{-1}$ (assuming CIE), which is $\sim 16$ 
percent of what is expected to be the contribution of the 
WHIM to $\Omega_b$ at $z=0$ and about 30 percent more
than what is estimated for intervening O\,{\sc vi} systems.
If we include the tentative detections we derive a 
high value for $\Omega_b$(BLA) of $0.0381\,h_{70}\,^{-1}$, suggesting
that a fair number of these tentative detections are not
related to hot gas in the WHIM but rather are caused by noise features
and non-thermal line broadening. 
These findings demonstrate 
that it is quite difficult to exactly pinpoint the baryon budget
of the BLAs with currently existing STIS data.
Next to the BLAs that are not related to the WHIM, the possible
overestimate of $\Omega_b$(BLA) due to non-thermal line 
broadening and, conversely, the possible underestimate of $\Omega_b$(BLA)
due to photoionization effects further complicate the situation.
However, we believe that the value of $\Omega_b$(BLA)$=0.0027\,h_{70}\,^{-1}$
is a reliable lower limit for the baryon content of BLAs
that lie above a detection limit of log $(N/b)\approx 11.3$.
Our study therefore suggests that broad Ly\,$\alpha$ 
absorbers arising in WHIM filaments 
contain a significant fraction of the baryons at $z=0$.

The BLAs represent an excellent way of studying the baryonic content of 
WHIM gas in the temperature range from $10^5$ to $10^6$ K. Although
there are substantial complications associated with the separation of 
the thermal and non-thermal broadening components and the 
uncertain role of photoionization, the fact that 
the baryon-content estimate does not require knowledge of 
the metallicity of the gas is a major advantage for using the
BLAs when determining the distribution and baryonic content of the WHIM.
A better understanding of these systems
will require additional high-resolution ultraviolet 
spectroscopic observations and detailed comparisons with models of 
the hydrodynamical evolution of the gas. Observations of the BLAs
with the Hubble Space Telescope {\it Cosmic Origins 
Spectrograph} (HST/COS) and future ultraviolet spectroscopic facilities 
may be the only direct way of probing low-metallicity regions in the 
cosmic web. Comparing the frequency of observed BLAs
with values of $dN/dz$ predicted by the simulations would help to constrain
the hot gas content of the web and test the importance of WHIM heat
sources other than gravitational collapse of the web filaments (i.e.,
galactic feedback by supernovae). 
Next to the important future X-ray observations of the WHIM
in emission and absorption,
spectral imaging of the Ly\,$\alpha$
emission over large enough fields of view to quantify the morphology 
of the WHIM gas would provide strong tests of the formation of 
large-scale gas structures and their relationship to galaxies. The WHIM
Ly\,$\alpha$ and metal-line emission is expected to be faint but could 
be observed with a moderate cost observatory designed for such an 
investigation (see Sembach et al.\,2005b for one such mission concept).

\begin{acknowledgements}

P.R. acknowledges financial support by the German
\emph{Deut\-sche For\-schungs\-ge\-mein\-schaft}, DFG,
through Emmy-Noether grant Ri 1124/3-1. B.D.S. acknowledges
financial support from NASA grant HST-GO-09184.03-A.

\end{acknowledgements}

{}


\begin{thebibliography}{}

\bibitem[]{}
Cen, R., \& Ostriker, J. 1999, ApJ, 514, 1

\bibitem[]{}
Chen, H.-W., \& Prochaska, J.X. 2000, ApJ, 543, L9

\bibitem[]{}
Danforth, C.W., \& Shull, J.M. 2005, ApJ, 624, 555

\bibitem[]{}
Dav\'e, R., Hernquist, L., Weinberg, D.H., \& Katz, N. 1997 ApJ, 477, 21 

\bibitem[]{}
Dav\'e, R., \& Tripp, T.M. 2001, ApJ, 553, 528

\bibitem[]{}
Dav\'e, R., Cen, R., Ostriker, J., et al.\,2001, ApJ, 552, 473

\bibitem[]{}
Fang, T., \& Bryan, G.L. 2001, ApJ, 561, L31

\bibitem[]{}
Fang, T., et al.\,2005, ApJ, 623, 612

\bibitem[]{}
Ferland, G.J., Korista, K.T., Verner, D.A., Fergueson, J.W., 
Kingdon, J.B., \& Verner, E.M. 1998, PASP, 110, 761

\bibitem[]{}
Fontana, A., \& Ballester, P. 1995, ESO Messenger, 80, 37

\bibitem[]{}
Fukugita, M. 2003, astro-ph 0312517

\bibitem[]{}
Kawahara, H., et al. 2005, astro-ph 0504594

\bibitem[]{}
Mathur, S., Weinberg, D.H., \& Chen, X. 2003, ApJ, 582, 82

\bibitem[]{}
Mazzotta, P., Mazzitelli, G., Colanfrancesco, S., \& Vittorio, N. 1998, 
A\&AS, 133, 403

\bibitem[]{}
Morton, D.C. 2003, ApJS 149, 205

\bibitem[]{}
Nicastro, F., et al.\,2005, Nature, 433, 495

\bibitem[]{}
Oegerle, W.R., et al.\,2000, ApJ, 538, L23

\bibitem[]{}
Penton, S.V., Stocke, J.T., \& Shull, J.M. 2004, ApJS, 152, 29

\bibitem[]{}
Richter, P., Savage, B.D., Tripp, T.M., \& Sembach, K.R. 2004, ApJS, 153, 165

\bibitem[]{}
Savage, B.D., Sembach, K.R., Tripp, T.M., \& Richter, P. 2002, ApJ, 564, 631

\bibitem[]{}
Savage, B.D., Lehner, N., Wakker, B.P., Sembach, K.R, \& Tripp, T.M. 2005, ApJ, 626, 776

\bibitem[]{}
Sembach, K.R., Tripp, T.M., Savage, B.D., \& Richter, P. 2004, ApJS, 155, 351

\bibitem[]{}
Sembach, K.R., et al.\,2005a, in preparation

\bibitem[]{}
Sembach, K.R., et al.\,2005b, "The Baryonic Structure Probe: Characterizing
the Cosmic Web of Matter Through Ultraviolet Spectroscopy", an Origins
Probe concept study submitted to NASA, May 13, 2005

\bibitem[]{}
Sutherland, R.S., \& Dopita, M.A. 1993, ApJS, 88, 253

\bibitem[]{}
Tripp, T.M., Lu, L., \& Savage, B.D. 1998, ApJ, 508, 200

\bibitem[]{}
Tripp, T.M., \& Savage, B.D. 2000, ApJ, 542, 42

\bibitem[]{}
Tripp, T.M., Savage, B.D., \& Jenkins, E.B. 2000, ApJ, 534, L1

\bibitem[]{}
Tripp, T.M., Giroux, M.L., Stocke, J.T., Tumlinson, J., \& Oegerle, W.R. 2001, ApJ, 563, 724

\bibitem[]{}
Tripp, T.M., et al.\,2005, in preparation

\bibitem[]{}
Valageas, P., Schaeffer, R., \& Silk  J. 2002, A\&A, 388, 741

\bibitem[]{}
Woodgate, B.E., et al.\,1998, PASP, 110, 1183

\end{thebibliography}
\end{document}